\documentclass[trsc,nonblindrev]{informs3} %

\OneAndAHalfSpacedXI %

\usepackage{amsmath}
\usepackage{color,soul}
\usepackage{algorithm}
\usepackage{algpseudocode}
\usepackage{bm}
\usepackage{multirow}

\usepackage{natbib}
 \bibpunct[, ]{(}{)}{,}{a}{}{,}%
 \def\bibfont{\small}%

\TheoremsNumberedThrough     %

\EquationsNumberedThrough    %

\graphicspath{ {./pictures/} }
\usepackage[justification=centering]{caption}
\usepackage{natbib}
\usepackage{subcaption}
\usepackage[skip=2pt,font=normalsize]{caption}

\begin{document}

\RUNAUTHOR{Chen et al.}

\RUNTITLE{A Multi-day Needs-based Modeling Approach for Activity and Travel Demand Analysis}

\TITLE{A Multi-day Needs-based Modeling Approach for Activity and Travel Demand Analysis}

\ARTICLEAUTHORS{%
\AUTHOR{Kexin (Sally) Chen}
\AFF{Massachusetts Institute of Technology, \EMAIL{kexinc@mit.edu}} %

\AUTHOR{Jinping (Jenna) Guan}
\AFF{Harbin Institute of Technology, \EMAIL{melon\_ping@163.com}}

\AUTHOR{Ravi Seshadri}
\AFF{Technical University of Denmark (DTU), \EMAIL{ravse@dtu.dk}}

\AUTHOR{Varun Pattabhiraman}
\AFF{Lyft, \EMAIL{varun89@gmail.com}}

\AUTHOR{Youssef Medhat Aboutaleb}
\AFF{Massachusetts Institute of Technology, \EMAIL{ymedhat@mit.edu}}

\AUTHOR{Ali Shamshiripour}
\AFF{The University of Arizona, \EMAIL{shamshiripour@arizona.edu}}

\AUTHOR{Chen Liang, Xiaochun Zhang}
\AFF{Shenzhen Urban Transport Planning Center Co., Ltd. (\EMAIL{liangchen@sutpc.com}, \EMAIL{lijunyuan@sutpc.com})}

\AUTHOR{Moshe Ben-Akiva}
\AFF{Massachusetts Institute of Technology, \EMAIL{mba@mit.edu}}

} %

\ABSTRACT{%
This paper proposes a multi-day needs-based model for activity and travel demand analysis. The model captures the multi-day dynamics in activity generation, which enables the modeling of activities with increased flexibility in time and space (e.g., e-commerce and remote working). As an enhancement to activity-based models, the proposed model captures the underlying decision-making process of activity generation by accounting for psychological needs as the drivers of activities. The level of need satisfaction is modeled as a psychological inventory, whose utility is optimized via decisions on activity participation, location, and duration. The utility includes both the benefit in the inventory gained and the cost in time, monetary expense as well as maintenance of safety stock. The model includes two sub-models, a {\it Deterministic Model} that optimizes the utility of the inventory, and an {\it Empirical Model} that accounts for heterogeneity and stochasticity. Numerical experiments are conducted to demonstrate model scalability. A maximum likelihood estimator is proposed, the properties of the log-likelihood function are examined and the recovery of true parameters is tested.
This research contributes to the literature on transportation demand models in the following three aspects. First, it is arguably better grounded in psychological theory than traditional models and allows the generation of activity patterns to be policy-sensitive (while avoiding the need for ad hoc utility definitions). Second, it contributes to the development of needs-based models with a non-myopic approach to model multi-day activity patterns. Third, it proposes a tractable model formulation via problem reformulation and computational enhancements, which allows for maximum likelihood parameter estimation.
}%


\maketitle

\setlength\parskip{1em plus 0.1em minus 0.2em}
\setlength\parindent{0pt}

\section{Introduction}

With recent developments in internet technologies, the flexibility to perform activities in time and space has grown extensively. The internet penetrates people's lives, allowing activities to be performed anywhere either online or offline. Such a shift in activity locations from physical to virtual has critical implications for transportation. For example, the impacts of e-commerce and remote working on transportation demand and thus emission is non-negligible \citep{greene_2023, international_transport_forum_2018, Le_et_al_2020}. Although we have now entered the post-pandemic era, fully remote or hybrid working may be permanent changes that can significantly affect daily transportation demand \citep{shreedhar_laffan_giurge_2022}. 

However, existing transportation demand models have several drawbacks that make them unsuitable for modeling the above trends. Specifically, the well-established rule-based and utility-based activity-based models (e.g., the day activity schedule approach, CEMDAP, FEATHERS, ADAPTS, and TASHA) often rely either directly on empirical data (i.e., sampling from empirical distributions) or on ad hoc definitions of utilities to model activity generation \citep{Bellemans_2010, AULD20121386, Bhat_2004, BOWMAN20011, miller_2003}. However, these approaches are increasingly insufficient given that activities have increasing flexibility, which has made it crucial to capture the underlying decision-making process of activity generation. On the other hand, there is an extensive literature on psychological theories that have described psychological needs as the propellers of human activities \citep{vansteenkiste_basic_2020}. Thus, a new modeling approach that incorporates these latent drivers (i.e., psychological needs) of activities needs to be developed to accurately represent activity and travel demand. 

This paper develops a multi-day needs-based model for transportation demand analysis. The fundamental concept underlying our approach posits that psychological needs motivate activities and activities influence the level of need satisfaction, which is modeled as a psychological inventory with its utility maximized. Activity decisions on participation, location, and duration are modeled, along with heterogeneity and stochasticity. This research advances the state-of-the-art in travel demand modeling in the following two aspects. {\bf First}, it advances activity-based modeling with a solid ground in psychological theories, which avoids several limitations of existing models: the empirically generated activity patterns and ad hoc utility formulation. {\bf Second}, it advances the existing needs-based models with a non-myopic approach for multi-day modeling. {\bf Third}, a tractable model formulation is developed by reformulating the utility maximization problem and incorporating computation enhancements, which enables maximum likelihood parameter estimation. 

In the following sections, a literature review is first presented in Section \ref{section:lit_review}, followed by the summary of the model framework in Section \ref{section:model_framework}. The formulation is presented in Section \ref{section:dm_formulation} and Section \ref{section:em_formulation}. The computation methods are summarized in Section \ref{section:computation}. Finally, Section \ref{section:numerical_experiment} shows the numerical experiment results and the paper concludes with Section \ref{section:conclusion}.

\section{Literature Review}\label{section:lit_review}

In this section, the activity-based models are first reviewed in Section \ref{section:abm_review}, followed by psychological theories in Section \ref{section:lit_review_psych_review}. Next, the existing needs-based models are reviewed in Section \ref{section:lit_review_need_based}. Finally, major contributions are summarized in Section \ref{section:contribution}. 

\subsection{Travel Demand Modeling: Activity-Based Model}\label{section:abm_review}

Modern transportation research requires disaggregate travel demand modeling. Advancements from early trip-based models to tour-based models have mitigated several limitations and captured inter-trip interactions \citep{BOWMAN20011}. The activity-based models further incorporate interactions among tours and are nowadays becoming more prevalent. 

Activity-based models consider travel demand as a derived demand of activities \citep{miller_2023}. Example activity-based model systems include the day activity schedule approach, CEMDAP, FEATHERS, ADAPTS, and TASHA \citep{AULD20121386, Bellemans_2010, Bhat_2004, BOWMAN20011, miller_2003}. The foundation was laid by Hägerstrand in 1970, whose time-geography theory suggests that demand for activity participation generates traveling behaviors to specific locations, at a certain time of day, and by particular travel modes \citep{Hagerstrand_1970}. Two types of activity-based models are often applied: the utility maximization-based econometric models and the rule-based computational process models \citep{de_luca_recent_2021}. The utility maximization-based models forecast activity and travel decisions with econometric models, including logit, probit, and ordered response models. On the other hand, the rule-based models impose heuristic condition-action rules and focus on scheduling and sequencing of daily activities. A detailed review of the activity-based models may be found in \cite{de_luca_recent_2021} and \cite{pattabhiraman_needs_based_2012}. Several advancements have been made to activity-based models recently. These include intra-household interactions, resource constraints (time use and budgeting), multi-day activity generation, activity planning and scheduling, and the inclusion of well-being and happiness indicators \citep{pattabhiraman_needs_based_2012}. 

Despite the wide applicability and richness of modeling capabilities, activity-based models have weaknesses yet to be addressed. {\bf First}, as mentioned by \cite{miller_2023} and \cite{pattabhiraman_needs_based_2012}, the generation of activity patterns and choice sets relies on observed patterns in the data. The mechanisms with which the patterns are formed, however, are not captured. {\bf Second}, for the utility maximization-based models, utilities are specified with various attributes and socio-demographics, but the construction of utility functions remains ad hoc and lacks support from behavioral theories. To address these limits, the underlying decision-making mechanisms need to be captured.

Thus, an alternative approach is necessary. Section \ref{section:lit_review_psych_review} reviews the literature on psychological theories on activity generation and develops the basis for the needs-based model that this research pursues. 

\subsection{Activity Generation: the Perspective of Psychologists}\label{section:lit_review_psych_review}

Psychologists have developed motivation theories to understand the psychological factors that drive human behavior. Two types of theories have been developed: the content theories that describe the triggering factors, and the process theories that focus on the connections between human behaviors and motivation \citep{rhee_2019}. 

\begin{figure}[htpb]
    \centering
    \includegraphics[width=4 in]{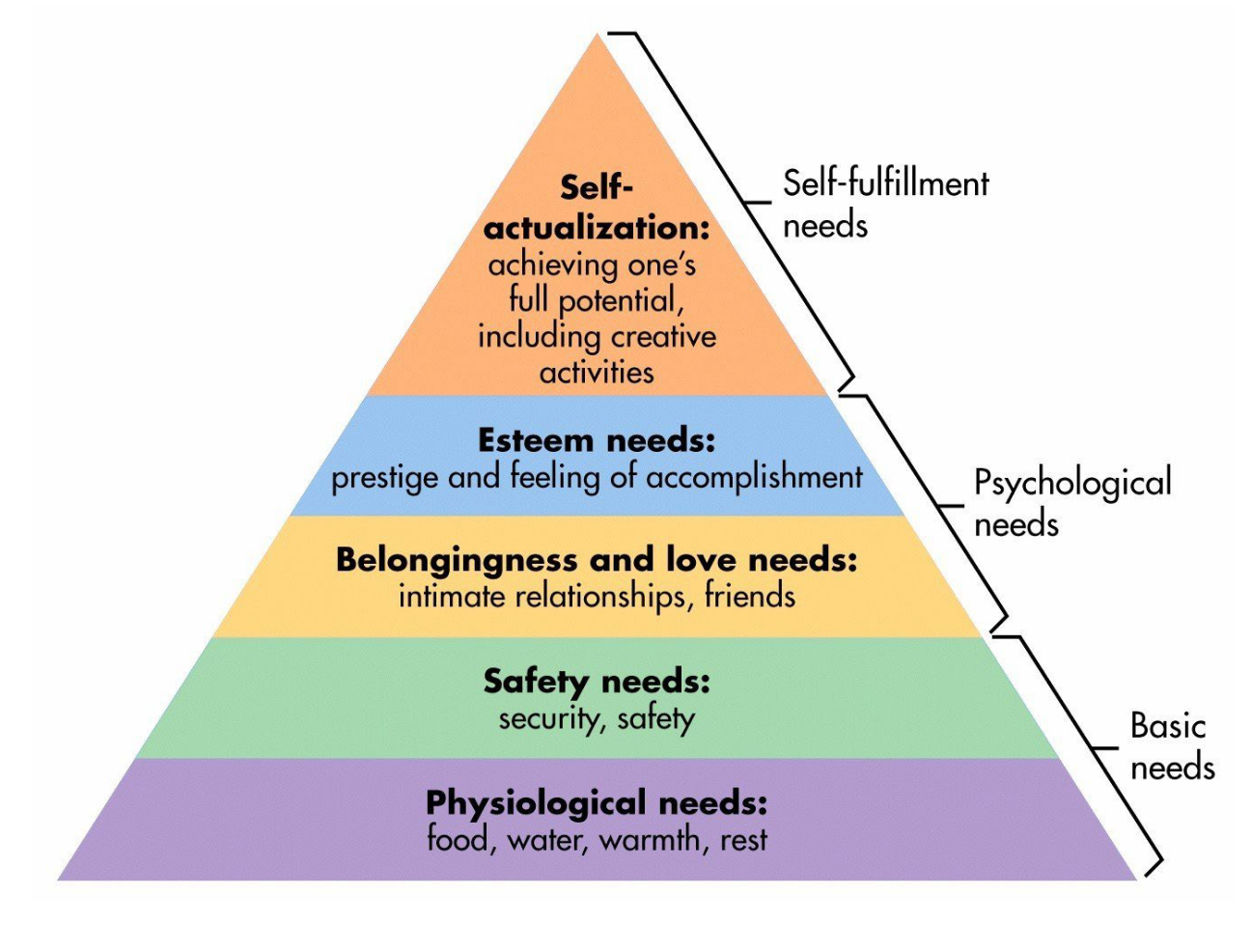}
    \caption{Maslow's Hierarchy of Needs \citep{mcleod_maslows_nodate}}
    \label{fig:maslow_pyramid}
\end{figure}

Maslow's Hierarchy of Needs is perhaps the most well-known content theory to describe human needs. The theory states that human needs are subject to a certain priority, and a five-level pyramid of needs has been developed, as shown in Figure~\ref{fig:maslow_pyramid}  \citep{maslov_1943, mcleod_maslows_nodate}. The hierarchy describes a decreasing trend in percentages of need satisfaction, and thus lower levels may not need to be fulfilled up to 100\% before proceeding to the higher levels \citep{maslow_1954}. Variations of Maslow's theory have been developed, including Alderfer’s ERG Theory which regroups needs into three categories: existence, relatedness, and growth needs \citep{aldefer_empirical_nodate}. In addition to the above two theories, a more general definition of need is adopted by the Basic Needs Theory: a motivating force is qualified as a need if it facilitates optimal psychological growth, integration, social development, and well-being \citep{ryan_self-determination_2000, ryan_overview_2002}. Herzberg’s Two Factor Theory is another renowned content theory that discusses the influencing factors of need growth and depletion \citep{johnson_irizarry_nguyen_maloney_2018}. Developed in the context of job satisfaction, Herzberg identified two classes of factors that motivate people to work: motivators (intrinsic) and hygiene factors (extrinsic) \citep{herzberg_1966}.

Psychologists have also developed process theories of motivation to explain the connection between behavior and motivations, as well as how motivation arises. The most relevant theory to this research is the Drive Reduction Theory, developed by psychologist Clark Hull. It is one of the major theories developed in the 1930s to depict the mechanism through which needs are influencing human behaviors \citep{hull_1943}. Per the Drive Reduction Theory, behaviors are functions of drive, which is defined as innate physiological needs, for example, food and water. When needs are deprived, tensions arise and thus actions are taken \citep{stults-kolehmainen_motivation_2020}.

In summary, abundant psychology research demonstrates a clear relationship between need and activity. As a result, it is necessary for transportation demand modelers to incorporate this crucial factor of need in model development, which has often been overlooked in the past.

\subsection{Needs-based Model}\label{section:lit_review_need_based}

\begin{figure}[htpb]
    \centering
    \includegraphics[width=3.5 in]{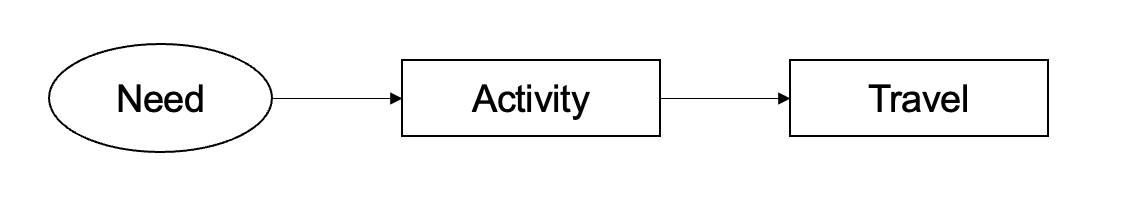}
    \caption{Relationships between Need, Activity and Travel}
    \label{fig:three_level_demand}
\end{figure}

Needs-based models have been developed to incorporate needs in the decision-making of activities, assuming activities are derived demand of latent needs, as shown in Figure \ref{fig:three_level_demand}. Among the existing studies, three approaches have been taken: sequential rule-based model, hazard-based duration model, and optimization-based model. 

The sequential rule-based model presents to be the first concrete step toward developing a needs-based model. Developed by \cite{arentze_2006}, \cite{arentze_need_based_2009}, and \cite{arentze_estimating_2011}, the model focuses on capturing intra-household interactions. The utility of activity is given by the level of need satisfaction. Decision rules are applied sequentially for each decision moment (e.g., hour or day) to determine the inclusion of activity, and need grows by a logistic curve automatically \citep{arentze_need_based_2009}. Multi-day data has been collected to empirically test the model \citep{nijland_representing_2013}. \cite{nijland_multi_day_2014} and \cite{nijland_incorporating_2012} extended the model by accounting for the impacts of pre-planned activities and events, and also tested it with empirical multi-day data. While these studies have contributed significantly by proposing an analytical framework for the needs-based model, some limitations remain unaddressed. {\bf First}, several critical activity dimensions are not considered, including the location. {\bf Second}, the models assume a sequential decision-making process,  which may fail to capture the multi-day dynamics and lead to myopic decisions. 

The hazard-based duration model, developed by \cite{cho_need-based_2023}, is another model that considers needs in modeling activity and travel decisions. Based on the needs-based theory proposed by \cite{arentze_2006}, a hazard-based duration approach is applied to model activity participation. The theory that the need for an activity increases over time is reflected in the positive slope of the hazard function. While the authors have demonstrated the model's applicability through empirical case studies, the model suffers a similar limitation to the rule-based models with the lack of activity dimensions to make decisions for. This includes the activity duration, which is an independent variable to predict frequency, but agents may make the decisions on duration and frequency simultaneously. 

The third group constructs the needs-based models for activity decisions as an optimization model. \cite{pattabhiraman_needs_based_2012} modeled the level of need satisfaction as a psychological inventory, and focuses on the prediction of the frequency, duration, expenditure, and location of activities. When an activity is conducted, the corresponding need is satisfied and the inventory is replenished. The optimization model maximizes the inventory. Monte Carlo experiment and empirical study were performed on a one-day travel diary data \citep{pattabhiraman_needs_based_2012}. 
A similar concept of modeling the need as an inventory was adopted by \cite{chow_multi-day_2015}. The formulated mixed integer linear program (MILP) optimizes activity scheduling, given the list of activities that need to be conducted for each day \citep{chow_multi-day_2015}. 
The optimization-based approach addresses several limitations of the sequential rule-based models and the hazard-based model. However, limitations still exist. While \cite{pattabhiraman_needs_based_2012} modeled for richer activity dimensions, it is a single-day model that is unable to capture multi-day dynamics. \cite{chow_multi-day_2015} focused on multi-day, but did not model the activity generation process, as well as the real-world heterogeneity and stochasticity.

\subsection{Contribution}\label{section:contribution}

This paper aims to develop a multi-day needs-based model to account for the underlying psychological drivers of activity generation. The level of need satisfaction is modeled as an inventory, similar to \cite{pattabhiraman_needs_based_2012} and \cite{chow_multi-day_2015}. The contribution of this research is threefold. 
{\bf First}, it is an extension of the activity-based model that is more solidly grounded in behavioral theories and thus addresses the shortcomings of the activity-based models, including the empirically generated activity patterns and ad hoc definition of utilities. {\bf Second}, via a non-myopic approach, the model mimics the multi-day decisions on multiple critical activity dimensions, including duration, location, and day participation. {\bf Third}, we propose a tractable model formulation by problem reformulation and incorporation of computation enhancements, and thus enables maximum likelihood parameter estimation that would otherwise be computationally intractable. 
Thus, this research advances the existing needs-based models with a more realistic approach to the modeling of activity decision-making processes.

\section{Model Framework}\label{section:model_framework}

\begin{figure}[htpb]
    \centering
    \includegraphics[width=2.7 in]{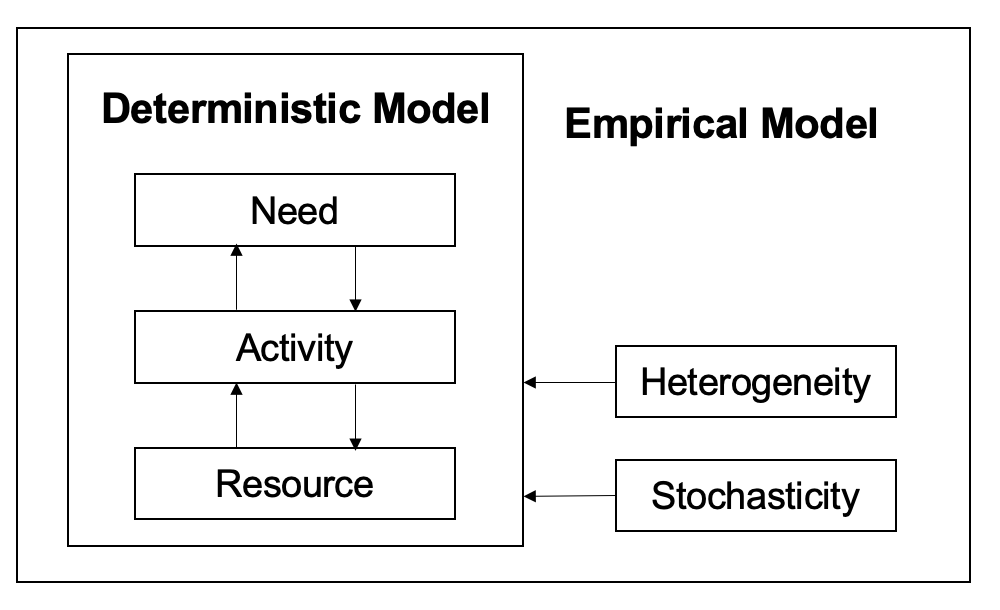}
    \caption{Needs-based Model Framework}
    \label{fig:needs_based_framework}
\end{figure}

The needs-based model developed in this research has two major components: the {\it Deterministic Model} and the {\it Empirical Model}, as shown in Figure~\ref{fig:needs_based_framework}. The prior models the relationships among three behavioral components: need, activity, and resource (i.e. materials necessary for performing activities). Decisions on activity participation, duration, and location are the outcomes of the {\it Deterministic Model}. The {\it Empirical Model} further incorporates stochasticity and heterogeneity in the real world and allows for parameter estimation of the needs-based model. 

To model the relationship among need, activity, and resource, the level of need satisfaction is represented as a psychological inventory, similar to the approach of \cite{pattabhiraman_needs_based_2012} and \cite{chow_multi-day_2015}. The inventory is replenished when activities are performed. When the inventory is low, the individual is motivated to perform an activity to replenish the inventory. Furthermore, the relation between need and activity may be complicated: i.e. one need may be satisfied by multiple activities and vice versa. We assume a one-to-one relation between need and activity: the need is satisfied by one particular activity and vice versa. 

In what follows, Section~\ref{dm_premise} first summarizes the behavioral premises considered. Section~\ref{dm_mechanism} details the mechanism of the psychological inventory.

\subsection{Behavioral Premises}\label{dm_premise}

Several behavioral premises are considered in the design of the {\it Deterministic Model}: 
\begin{enumerate}
    \item {\it Growth of need}. Need increases over time when the activities are not conducted, which is consistent with the cyclic pattern of activities. The day of the week may impact the rate at which the need grows. Such a construction of a need with both automatic growth and impacts from external factors is consistent with Herzberg’s Two Factor Theory \citep{herzberg_1966}. 
    
    \item {\it Factors of need satisfaction}. We consider two factors of need satisfaction: contextual factors and human behaviors. Contextual factors include environmental and social factors, e.g. attractiveness of the location, weather, social network quality, and culture \citep{abidin_social_2021}. Based on theories of motivation, human behaviors, e.g., decisions on activity duration, are motivated by need and thus impact the level of need satisfaction \citep{hull_1943, stults-kolehmainen_motivation_2020}. 
    
    \item {\it The Rational Choice Theory}. Rational agents evaluate costs and benefits when making decisions \citep{browning_understanding_2000}. The benefits include the psychological inventory gained by performing activities, and the costs include the disutilities of time, maintenance of safety stock, and travel costs incurred to perform the activity. 

    \item {\it Basic Needs Theory}. The utility of the psychological inventory is maximized, which is consistent with the qualification of need imposed by \cite{ryan_self-determination_2000} who states that need must facilitate optimal psychological growth, integration, social development, and well-being.

    \item {\it Resource constraint}. Resources are limited in activity planning. The limitation in daily free time is thus considered, which is defined as the available time allocated to perform the replenishment activity. 
\end{enumerate}

\subsection{Mechanism}\label{dm_mechanism}
Based on the behavioral premises, the mechanism of the psychological inventory to capture the relationships among the need, activity, and resource is now presented. For each day, an activity may be performed to increase the psychological inventory. This increase will be hereafter referred to as the activity production and is dependent on location attractiveness and the activity duration. When an activity is performed, the inventory is replenished at the beginning of the day. Furthermore, regardless of activity participation, the inventory depletes over time at a given rate. Four constraints are imposed:
\begin{enumerate}
    \item {\it Conservation of psychological inventory}. The inventory at the beginning of a day is equal to the inventory at the beginning of the previous day, plus the replenished amount, and minus the depleted amount. 
    \item {\it Periodicity}. The inventory at the end of the planning horizon is equal to the amount at the beginning of the first day, assuming a balanced inventory over time. 
    \item {\it Replenish condition}. The amount of inventory of a day, after the replenishment activity (if any), is able to support the depletion over the same day. 
    \item {\it Daily time constraint}. Time spent on an activity and related travel is less than the free time allocated to perform the activity. 
\end{enumerate}

We assume that individuals plan for activities to maximize the overall utility throughout the planning horizon. Utility measures the trade-off between cost and benefit. Having a positive amount of psychological inventory brings benefits and increases the utility. On the other hand, maintenance of a safety stock, time, and money spent are costs that bring disutility. 

\begin{figure}[htpb]
    \centering
    \includegraphics[width=3 in]{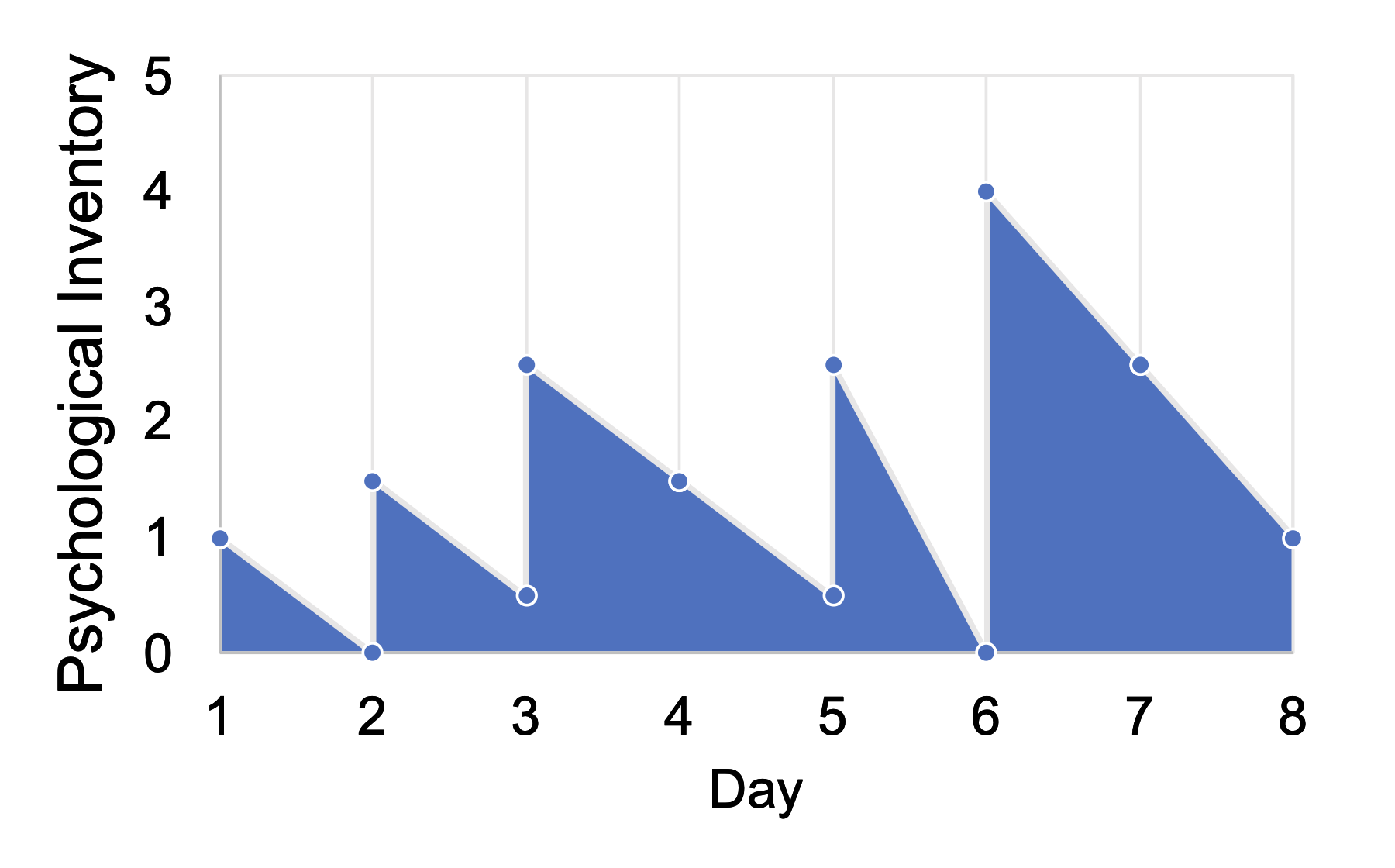}
    \caption{Psychological Inventory Illustration}
    \label{pi_illustration}
\end{figure}

Figure~\ref{pi_illustration} illustrates a psychological inventory over a one-week planning horizon. Activities are performed on day 2, 3, 5, and 6. All constraints are satisfied. 

\section{Model Formulation: Deterministic Model}\label{section:dm_formulation}

The formulation of the {\it Deterministic Model} is now presented. The notation is first introduced in Section \ref{section:dm_notation}, followed by the formulation in Section \ref{section:dm_opt_formulation}. The derivation of a tractable formulation is presented in Section \ref{section:dm_reformulation}.

\subsection{Notation}\label{section:dm_notation}

Consider a planning horizon of length $H$ days. The following section describes the notations used for the {\it Deterministic Model}.

{\bf Decision Variables}. Three decision variables are considered for modeling the multi-day activity pattern: activity participation ($\{\delta_t\}_{t=1}^H)$, duration ($\{d_t\}_{t=1}^H)$,  location ($\{i_t\}_{t=1}^H$). Participation $\delta_t$ is binary, and $\delta_t=1$ indicates that the activity is performed on day $t$. Duration $d_t$ represents the time spent on the activity (excluding travel time) on day $t$. When the activity is not performed ($\delta_t=0$), duration is $0$, otherwise positive . The third decision variable $i_t$ denotes the location to perform the activity on day $t$, and belongs to the set of locations ($\mathcal{L}$). 

{\bf Attributes}. For location $j\in \mathcal{L}$, day $t\in\{1,\ldots,H\}$, three attributes are considered: location attractiveness ($A_{j,t}$), two-way travel time between home and the location ($TT_{j,t}$), and two-way travel cost between home and the location ($TC_{j,t}$). Let $\bm{A}$, $\bm{TT}$, and $\bm{TC}$ denote the corresponding matrices. 

{\bf Individual Characteristics}. For individual characteristics, four items are considered: free time on day $t$ ($FT_t$), consumption rate on day $t$ ($\lambda_t$), value of time ($\rho_1$), value of safety stock ($\rho_2$), and value of inventory ($\rho_3$). 
Free time imposes a constraint on the time available for performing the activity. The free time over the planning horizon is represented by a vector of length $H$: $\bm{FT}=[FT_1, \ldots, FT_H]^T$. 
The consumption rate represents the amount of inventory depleted over the day. In general, the consumption rate per day may vary across days, composing a vector of rates $\bm{\lambda}=[\lambda_1,\ldots, \lambda_H]^T$. For simplicity, we assume a rate of $\lambda$ for all weekdays and $\gamma\lambda$ for all weekend days, thus capturing the effect of the day of the week. 
The latter three characteristics, $\rho_1$, $\rho_2$ and $\rho_3$, respectively represent the monetary value of time, safety stock (i.e., minimum $I_t$ over the planning horizon), and psychological inventory.

{\bf The Psychological Inventory}. For each day $t\in \{1,\ldots,H\}$, we define the following two quantities: psychological inventory at the beginning of the day ($I_t$) and activity production ($Q_t$). We assume that if the activity is performed on day $t$, the production it generates will be replenished at the beginning of the day. The activity production $Q_t$ is a non-negative function that is dependent on decisions regarding activity participation ($\delta_t$), duration ($d_t$), and location ($i_t$). Similar to \cite{pattabhiraman_needs_based_2012}, the commonly used Cobb-Douglas production function is used. Three parameters are included: a constant ($q_0$), elasticity of activity duration ($q_1$), and elasticity of location attractiveness ($q_2$). As activity duration $d_t=0$ when $\delta_t=0$, $Q_t$ is conditional on $\delta_t$:
\begin{align}
    Q(d_t, i_t) = \begin{cases}
    \exp \left[q_0 + q_1 \ln (d_t) + q_2 \ln (A_{i_t, t})\right], & \text{if } \delta_t = 1\\
    0,              & \text{if } \delta_t=0
        \end{cases} \label{eq:dm_prod_def}
\end{align}

{\bf Objective Function}. The objective is to maximize the utility of the inventory over the planning horizon $H$, which includes a trade-off between benefit and cost. The benefit is the average psychological inventory per day over the planning horizon, which is the average of the total shaded area in Figure~\ref{pi_illustration}. 
The cost is composed of three components: average time spent on the activity and travel ($\frac{1}{H}\ \sum_{t=1}^H (d_t + \delta_t TT_{i_t,t})$), safety stock that is defined as the minimal psychological inventory across all days ($I_{min}$), and average travel cost ($\frac{1}{H}\ \sum_{t=1}^H (\delta_t TC_{i_t, t}$). 
The parameters $\rho_1$, $\rho_2$, and $\rho_3$ translate the cost components in the utility into monetary units.

\subsection{Optimization Problem Formulation}\label{section:dm_opt_formulation}
With the above-mentioned definitions, the {\it Deterministic Model} to optimize an individual's inventory is formulated in Equations (\ref{eq:dm_obj}) to (\ref{eq:dm_i_value}). 
Equation (\ref{eq:dm_obj}) defines the average utility of an individual over the planning horizon. Equation (\ref{eq:dm_ss_def}) defines the safety stock. Equation (\ref{eq:dm_cons}), (\ref{eq:dm_periodicity}), (\ref{eq:dm_rep_cond}) and (\ref{eq:dm_time_cons}) respectively model the constraints of {\it Conservation of psychological inventory}, {\it Periodicity}, {\it Replenish condition}, and {\it Daily time constraint} as described in Section~\ref{dm_mechanism}.
\begin{align}
    DM (H |\bm{\lambda}, \bm{FT}, &\bm{A}, \bm{TT}, \bm{TC}, q_0, q_1, q_2, \rho_1, \rho_2, \rho_3) = & & \nonumber \\
    \max_{\{\delta_t\},\{d_t\},\{i_t\}} & \ \frac{\rho_3}{H}\ \sum_{t=1}^H \left[I_t+Q(d_t, i_t) -\frac{\lambda_t}{2}\right] & \nonumber \\
    & - \left[\frac{\rho_1}{H}\ \sum_{t=1}^H (d_t + \delta_t TT_{i_t,t}) + \rho_2 I_{min} + \frac{1}{H}\ \sum_{t=1}^H (\delta_t TC_{i_t, t}) \right] & \label{eq:dm_obj}\\
    \text{s.t.} \quad\quad & I_{min} = \min_{t\in\{1,\ldots,H\}} I_t & & \label{eq:dm_ss_def}\\
            & I_{t+1} = I_t + Q_t - \lambda_t &\quad\forall t\in\{1,\ldots,H-1\} &\label{eq:dm_cons}\\
            & I_1 = I_H + Q_H - \lambda_H & & \label{eq:dm_periodicity}\\
            & I_t + Q_t \geq \lambda_t    &\quad\forall t\in\{1,\ldots,H\} &\label{eq:dm_rep_cond}\\
            & d_t + \delta_t TT_{i_t,t} \leq FT_{t}     &\quad\forall t\in\{1,\ldots,H\} &\label{eq:dm_time_cons}\\
            & \delta_t \in \{0, 1\}                     &\quad\forall t\in\{1,\ldots,H\} &\label{eq:dm_delta_value}\\
            & d_t \begin{cases}
                    > 0, & \text{if } \delta_t = 1\\
                    = 0,   & \text{if } \delta_t=0 \end{cases} & \quad\forall t\in\{1,\ldots,H\} & \label{eq:dm_d_value}\\
            & i_t \in \mathcal{L} & &\label{eq:dm_i_value}
    \stepcounter{equation}
\end{align}

For various activities (e.g., grocery shopping, social activity), one week is a proper planning horizon. It is possible, however, that the optimal objective value is negative. This indicates that performing the activity on a weekly basis leads to disutility, which suggests the necessity of having less frequent but more efficient replenishment activities. Therefore, a longer planning horizon is needed, while the inputs are the same among different weeks. Algorithm (\ref{alg:multi_week_model}) may thus be used for multi-week modeling.
\begin{algorithm}
\caption{Algorithm for Multi-week Modeling}\label{alg:multi_week_model}
\begin{algorithmic}
\State $H \gets 7$
\State $V^* \gets DM (H |\bm{\lambda}, \bm{FT}, \bm{A}, \bm{TT}, \bm{TC}, q_0, q_1, q_2, \rho_1, \rho_2, \rho_3)$
\While{$V^* < 0$}
    \State $H \gets H+7$
    \State $\bm{\lambda} \gets \bm{\lambda}.append\left(\bm{\lambda}[1:7]\right)$
    \State $\bm{FT} \gets \bm{FT}.append\left(\bm{FT}[1:7]\right)$
    \State $V^* \gets DM (H |\bm{\lambda}, \bm{FT}, \bm{A}, \bm{TT}, \bm{TC}, q_0, q_1, q_2, \rho_1, \rho_2, \rho_3)$
\EndWhile
\end{algorithmic}
\end{algorithm}

\subsection{Optimization Problem Reformulation}\label{section:dm_reformulation}
The {\it Deterministic Model} is reformulated to a tractable formulation using three strategies. First, lifting is applied by using auxiliary variables to lift the problem to a higher space where it is linear. Second, logical conditions are formulated using the big M approach. Third, where necessary, constraints are transformed into convex cones (i.e., Equations (\ref{eq:cost_of_time_with_eta}) to (\ref{eq:eta_power_cone})). The following section summarizes the reformulation of the model into a mixed integer conic program (MICP). 

\subsubsection{Reformulation of Objective Function}

{\bf Production Function}. 
The production function defined by Equation (\ref{eq:dm_prod_def}) is first transformed into a linear function via lifting. Let $x_{jt}\in \{0, 1\}$ indicate whether an activity is performed at location $j\in \mathcal{L}$ on day $t$ or not. Then, $Q(d_t, i_t)$ is equivalent to:
\begin{align}
    Q(d_t, i_t) &= \sum_{j\in \mathcal{L}}{x_{jt}\exp{\left[q_0+q_1\ln{\left(d_t\right)}+q_2\ln{\left(A_{j,t}\right)}\right]}} = d_t^{q_1\ }\sum_{j}{x_{jt}e^{q_0}A_{j,t}^{q_2}} 
\end{align}

Define a new continuous variable $\tau_t = d_t^{q_1}$, and hence:
\begin{align}
    d_t=\tau_t^{q_1^{-1}} \label{eq:dm_reform_d_as_tau}
\end{align} 

The production function can be now formulated as Equations (\ref{eq:dm_refom_Q}) to (\ref{eq:dm_refom_tau_lower0}), with additional logical constraints, where $M_0$ is an arbitrarily large number, $M_1$ imposes an upper bound on $\tau_t$ and $y_{jt}$ is binary. For $t\in\{1,\ldots,H\}$: 
\begin{gather}
    Q\left(d_t,i_t\right) =\tau_t\sum_{j\in \mathcal{L}}{x_{jt}e^{q_0}A_{j,t}^{q_2}} = \sum_{j\in \mathcal{L}}{y_{jt}e^{q_0}A_{j,t}^{q_2}} \label{eq:dm_refom_Q} \\ 
    y_{jt} \le\ M_0 x_{jt} \\ 
    y_{jt} \le\tau_t  \\ 
    y_{jt} \geq\tau_t-\left(1-x_{jt}\right)M_0  \\ 
    y_{jt} \geq 0 \\
    \sum_{j\in \mathcal{L}}{x_{jt}=\delta_t}  \label{eq:dm_refom_delta_as_x}  \\
    \tau_t\le\ M_1\delta_t  \label{eq:dm_refom_tau_upperM} \\ 
    \tau_t\geq0  \label{eq:dm_refom_tau_lower0}
\end{gather}

{\bf Cost of Time}. 
By Equation (\ref{eq:dm_reform_d_as_tau}), the cost of time has a complex component ($\tau_t^{q_1^{-1}}$). Introduce a new variable $\eta_t$ to rewrite the cost of time term in Equation (\ref{eq:dm_obj}): 
\begin{gather}
    \frac{\rho_1}{H}\sum_{t=1}^{H}\left(\eta_t+\sum_{j}{x_{jt}TT_{j,t}}\right) \label{eq:cost_of_time_with_eta} \\
    \tau_t^{q_1^{-1}}\le\eta_t \label{eq:eta_constraint}
\end{gather}

Equation (\ref{eq:eta_constraint}) may be written as a 3D power cone:
\begin{align}
    \tau_t\le\eta_t^{q_1}\cdot1^{1-q_1} \quad\Leftrightarrow\quad \left(\eta_t,1,\tau_t\right)\in\mathcal{K}_{q_1} \label{eq:eta_power_cone}
\end{align}

{\bf Safety Stock}. 
Logical constraints are imposed to compute the safety stock ($I_{min}$). Define binary variables $w_t$ ($\forall t\in\{1,\ldots,H\}$). Then, the following linear constraints enforce $I_{min}$ to be the minimum inventory, where $M_2$ is an arbitrarily large number: 
\begin{align}
    &I_{min} \le\ I_t,   &\forall\ t\in \{1,\ldots,H\} \\
    &I_{min} \geq\ I_t-M_2 w_t, &\forall\ t\in \{1,\ldots,H\} \\
    &\sum_{t=1}^{H}w_t=H-1 
\end{align}

{\bf Monetary Expense}. 
By Equation (\ref{eq:dm_refom_delta_as_x}), the $\delta_t$ variable is now replaced by $x_{jt}$, and therefore: 
\begin{align}
    \frac{1}{H}\sum_{t=1}^{H}\left(\delta_t{TC}_{i_t,t}\right) = \frac{1}{H}\sum_{t=1}^{H} \sum_{j\in \mathcal{L}}{x_{jt}TC_{j,t}}  
\end{align}

\subsubsection{Reformulation of Constraints}

The production function, $Q$, is utilized in the definition of the constraints of {\it Conservation of psychological inventory}, {\it Periodicity} and {\it Replenish condition}, respectively Equation (\ref{eq:dm_cons}), (\ref{eq:dm_periodicity}) and (\ref{eq:dm_rep_cond}). By Equation (\ref{eq:dm_refom_Q}), the above three constraints are reformulated to:
\begin{align}
    &I_{t+1} = I_t + \sum_{j\in \mathcal{L}}{y_{jt}e^{q_0}A_{j,t}^{q_2}} - \lambda_t &\quad\forall t\in\{1,...,H-1\}\\
    &I_1 = I_H + \sum_{j\in \mathcal{L}}{y_{jH}e^{q_0}A_{j,H}^{q_2}} - \lambda_H &  \\
    &I_t + \sum_{j\in \mathcal{L}}{y_{jt}e^{q_0}A_{j,t}^{q_2}} \geq \lambda_t    &\quad\forall t\in\{1,...,H\} 
\end{align}

Finally, consider the {\it Daily time constraint} as given by Equation (\ref{eq:dm_time_cons}). By Equation (\ref{eq:dm_reform_d_as_tau}) and (\ref{eq:dm_refom_delta_as_x}), Equation (\ref{eq:dm_time_cons}) is equivalent to:
\begin{align}
    \tau_t^{q_1^{-1}}+\sum_{j\in \mathcal{L}}\ x_{jt}TT_{jt}\le\ FT_t\ \quad\Leftrightarrow\quad \tau_t\le\left(FT_t-\sum_{j\in \mathcal{L}}\ x_{jt}TT_{jt}\right)^{q_1} \quad \forall t\in\{1,\ldots,H\}  
\end{align}

By Equation (\ref{eq:dm_refom_delta_as_x}) to (\ref{eq:dm_refom_tau_lower0}), $\tau_t=0$ when $x_{jt}=0 \ \forall j,t$. Hence, the constraint may be transformed into: 
\begin{align}
    \tau_t\le\sum_{j\in \mathcal{L}}\ x_{jt}\left(FT_t-TT_{jt}\right)^{q_1} \quad \forall t\in\{1,\ldots,H\}
\end{align}

As a result, the reformulated model is a MICP, which can be solved using existing solvers (e.g., Mosek) \citep{mosek}. %

\section{Model Formulation: Empirical Model}\label{section:em_formulation}

Building on the {\it Deterministic Model}, the {\it Empirical Model} incorporates heterogeneity and stochasticity. For simplicity, we present the model for $H=7$ but the model can be generalized for an arbitrary length of the planning horizon. 

Consider a sample of $N$ individuals. For each individual $n \in \{1,...,N\}$, we observe the activity pattern over the planning horizon, including activity participation, duration, and location. 
Denote the observed pattern by three vectors of length $H=7$: $\bm{\delta_n}=[\delta_{1,n},...,\delta_{7,n}]^T$, $\bm{d_n}=[d_{1,n},...,d_{7,n}]^T$, and $\bm{i_n}=[i_{1,n},...,i_{7,n}]^T$. 
At the same time, the {\it Deterministic Model} computes an optimal solution ($\bm{\delta_n^*}$, $\bm{d_n^*}$, $\bm{i_n^*}$) of length $H^*$, given certain parameters and inputs:
 \begin{enumerate}
     \item The parameters include both random ($\bm{\zeta_n}\sim D(\bm{\Theta})$) and non-random parameters ($\bm{\xi}$). The random parameters follow a distribution of $D(\bm{\Theta})$, where $D$ denotes an arbitrary distribution parameterized by $\bm{\Theta}$. The specific parameters included in $\bm{\zeta_n}$ and $\bm{\xi}$ vectors are summarized in Section~\ref{section:em_input_param}. 
     
     \item The inputs for individual $n$ are $\bm{x_n}=[\bm{FT}_n, \bm{TT}, \bm{TC}, \bm{A}]^T$, where $\bm{FT_n}$ is a vector of free time of length 7, and $\bm{TT}, \bm{TC}, \bm{A}$ respectively are the travel time matrix, travel cost matrix, and vector of location attractiveness. 
 \end{enumerate}

Due to stochasticity and heterogeneity, the observed pattern and the optimal pattern may not match, including the length ($H$ vs. $H^*$). The value of $H^*$ may be higher than $H$, given Algorithm (\ref{alg:multi_week_model}). Let $H^* = 7K^*,\ K^*\in\mathbb{Z}^+$. The number of weeks solved by the {\it Deterministic Model} is hence represented by $K^*$ and may be greater than one. The {\it Empirical Model} accounts for this difference in length. In what follows, Section~\ref{section:em_prob_model} presents the probabilistic model, and Section \ref{section:em_input_param} summarizes the inputs and parameters. Section~\ref{section:em_likelihood} describes the likelihood function. Finally, Section~\ref{section:em_estimation} discusses the estimation methods. 

\subsection{Probabilistic Model}\label{section:em_prob_model}

The probabilistic model computes the joint probability of the observed location, duration, and participation, conditional on inputs, random and non-random parameters. It involves a joint probability function of location and participation choice, and a probability density function for the duration, as shown in Equation (\ref{eq:prob_decompose}). The following section presents the formulation of the two probabilistic models.
\begin{align}
    P(\bm{i_n}, \bm{\delta_n}, \bm{d_n} | \bm{\zeta_n}, \bm{\xi}, \bm{x_n}) = P(\bm{i_n}, \bm{\delta_n} | \bm{\zeta_n}, \bm{\xi}, \bm{x_n}) \times f(\bm{d_n} | \bm{i_n}, \bm{\delta_n}, \bm{\zeta_n}, \bm{\xi}, \bm{x_n})
    \label{eq:prob_decompose}
\end{align}

\subsubsection{Joint Location and Participation Choice}

\begin{figure}[htpb]
    \centering
    \includegraphics[width=2.5 in]{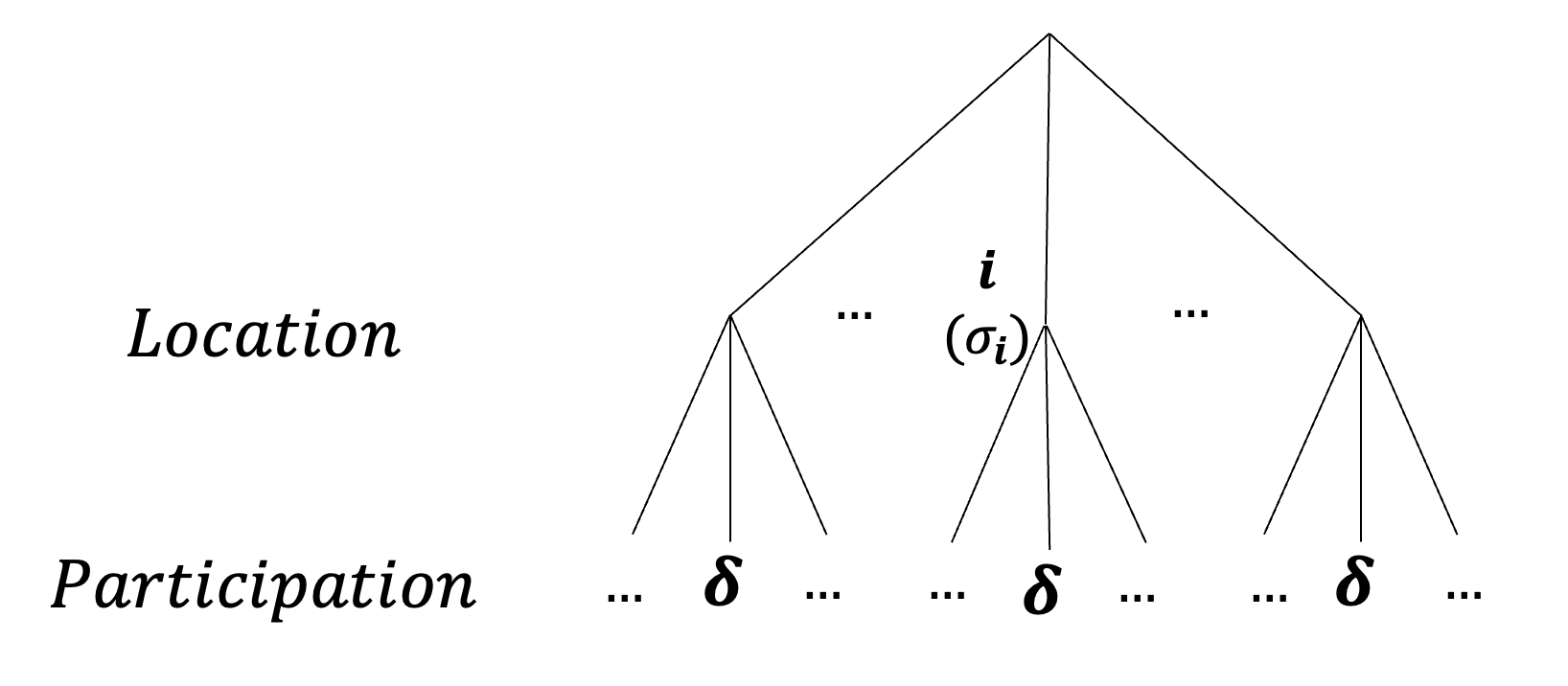}
    \caption{Nesting Structure for Joint Choice of Location and Participation}
    \label{fig:nest_loc_part}
\end{figure}

The joint probability of location and participation is modeled by a logit mixture model with an error component to capture the nesting structure shown in Figure \ref{fig:nest_loc_part}. The top level consists of all alternatives of locations, and the lower level consists of all alternatives of participation. Each nest $\bm{i}$ is associated with a normally distributed error component $\eta_{\bm{i}}\sim N(0, \sigma_{\bm{i}}^2)$. 

Denote the set of all location and participation choices as $\Delta$. The systematic utility of an alternative $(\bm{i}, \bm{\delta}) \in \Delta$ is computed as a summation of three components:
\begin{enumerate}
    \item Optimal utility, computed from {\it Deterministic Model}. Given the parameters $\bm{\zeta_n}$, $\bm{\xi}$, and input $\bm{x_n}$, by fixing $\bm{i}$ and $\bm{\delta}$, the {\it Deterministic Model} solves for an optimal solution with the restricted activity pattern. Hence, among the three activity dimensions of location, participation, and duration, only duration is optimized by the {\it Deterministic Model}. Denote the corresponding optimal objective value as $\Tilde{V}(\bm{i}, \bm{\delta} | \bm{\zeta_n}, \bm{\xi}, \bm{x_n})$.
    
    \item Size measures. Denoted as $M_{\bm{i}}$, it evaluates the size of the location $\bm{i}$ (see \cite{Ben-Akiva_Lerman_2018}), and is computed as a linear function of measures of size of location $\bm{i}$ as shown in Equation (\ref{eq:size_measure}), where $k$ denotes a measure of size (e.g., employment size and zone area) and $\bm{\beta}$ denotes associated parameters. Note that since $\bm{i}$ is a vector that may be composed of multiple locations, $x_{\bm{i}, k}$ is defined here as a generic measure, and may be constructed via various means. In the case when $\bm{i}$ only consists of one unique location, $x_{\bm{i}, k}$ may directly be the measured value. 
        \begin{align}
            M_{\bm{i}} = \sum_k \beta_k x_{\bm{i}, k} \label{eq:size_measure} 
        \end{align}
    \item Error component. The error component $\eta_{\bm{i}}\sim N(0, \sigma_{\bm{i}}^2)$ captures the correlation among alternatives within the same nest. 
\end{enumerate}

Therefore, the systematic utility of a joint location and participation alternative is computed by Equation (\ref{eq:loc_part_sys_util}). Equation (\ref{eq:loc_part_util}) presents the utility with an extreme value distributed error. The conditional probability of observing location $\bm{i_n}$ and participation $\bm{\delta_n}$ is given by Equation (\ref{eq:loc_part_prob}), where $\mu$ is the scale parameter:
\begin{align}
    &V(\bm{i}, \bm{\delta} | \bm{\zeta_n}, \bm{\xi}, \bm{x_n}) = \Tilde{V}(\bm{i}, \bm{\delta} | \bm{\zeta_n}, \bm{\xi}, \bm{x_n}) + \ln(M_{\bm{i}}) + \eta_{\bm{i}} \label{eq:loc_part_sys_util} \\ 
    &U(\bm{i}, \bm{\delta} | \bm{\zeta_n}, \bm{\xi}, \bm{x_n}) = V(\bm{i}, \bm{\delta} | \bm{\zeta_n}, \bm{\xi}, \bm{x_n}) + \varepsilon_{\bm{i}, \bm{\delta}, n}, \quad\varepsilon_{\bm{i}, \bm{\delta}, n}\sim \text{Extreme Value Type I}(0, \mu) \label{eq:loc_part_util} \\
    &P(\bm{i_n}, \bm{\delta_n} | \bm{\zeta_n}, \bm{\xi}, \bm{x_n}) = \frac{\exp[\mu\times V(\bm{i_n}, \bm{\delta_n} | \bm{\zeta_n}, \bm{\xi}, \bm{x_n})]}{\sum_{\{\bm{i}, \bm{\delta}\} \in \Delta} \exp[\mu\times  V(\bm{i}, \bm{\delta} | \bm{\zeta_n}, \bm{\xi}, \bm{x_n})]}\label{eq:loc_part_prob}
\end{align}

\subsubsection{Duration Probability}
Similar to \cite{pattabhiraman_needs_based_2012}, we assume measurement errors in the observed duration. Conditioned on the observed location and participation, the optimal duration can be computed using the {\it Deterministic Model}. 
Denote the optimal duration vector as $\bm{d}_{\bm{i_n},\bm{\delta_n},n}^*=d^*(\bm{i_n},\bm{\delta_n}|\bm{\zeta_n},\bm{\xi},\bm{x_n})$, with length $H^*=7K^*$ that may exceed $H$.
To account for this discrepancy, the duration probability is constructed as an average probability of observing $\bm{\delta_n}$ in each of the $K^*$ weeks, as shown in Equation (\ref{eq:dur_prob}). The function $G$ computes the probability of observing $\bm{d_n}$ at week $k$, conditioned on location $\bm{i_n}$ and participation $\bm{\delta_n}$.
\begin{align} 
    f(\bm{d_n} |\bm{i_n},\bm{\delta_n},\bm{\zeta_n},\bm{\xi},\bm{x_n})=\frac{1}{K^*} \sum_{k=1,...,K^*} G\left(\bm{d_n},k|\bm{i_n},\bm{\delta_n},\bm{\zeta_n},\bm{\xi},\bm{x_n} \right) \label{eq:dur_prob}
\end{align}

To derive $G$, consider a week $k\in \{1,...,K^*\}$ solved by the {\it Deterministic Model}. 
Suppose the optimal duration of the week $k$ is $[d_{1,n}^{*k}, \ldots, d_{7,n}^{*k}]^T$ with length 7. 
For each day $t\in\{1,...,7\}$, consider a multiplicative measurement error $\nu_{t,n}^k$ in the observed duration, as expressed by Equation (\ref{eq:dur_measurement_err}). We assume the measurement error is independent across days. 
The probability density function of observing duration $d_{t,n}$ from week $k$ can then be derived as Equation (\ref{eq:dur_prob_one_day_from_one_week}), 
where $\phi()$ here is the probability density function of a standard normal random variable.
\begin{align}
    &d_{t,n}=d_{t,n}^{* k} \exp(\nu_{t,n}^k),\ \nu_{t,n}^k\sim N(0,\sigma_{dur}^2\ ) \label{eq:dur_measurement_err} \\
    &g_{d_{t,n}}^k = g\left(d_{t,n},k | \bm{i_n},\bm{\delta_n},\bm{\zeta_n},\bm{\xi},\bm{x_n}\right)=\frac{1}{d_{t,n} \sigma_{dur}} \phi\left[\frac{\ln(d_{t,n})-\ln(d_{t,n}^{* k})}{\sigma_{dur}} \right] \label{eq:dur_prob_one_day_from_one_week}
\end{align}

Finally, the probability of observing the duration $\bm{d_n}$ from week $k$ is derived as:
\begin{align}
    G(\bm{d_n},k|\bm{i_n},\bm{\delta_n},\bm{\zeta_n},\bm{\xi},\bm{x_n}) = \prod_{t\in\{1,..,7\}}\left(g_{d_{t,n}}^k\right)^{\delta_{t,n}} \label{eq:dur_G}
\end{align}

\subsection{Inputs and Parameters}\label{section:em_input_param}

The needs-based model predicts the activity dimensions as the dependent variables, namely $\bm{\delta_n}$, $\bm{d_n}$, and $\bm{i_n}$. The independent variables are $\bm{x_n}=[\bm{FT}_n, \bm{TT}, \bm{TC}, \bm{A}]^T$. Free times are assumed to be the same as $FT_{WD,n}$ for weekdays and the same as $FT_{WE,n}$ for weekend days.

The non-random parameters in the model are $\bm{\xi}=[\lambda, \gamma, q_1, q_2, \mu, \bm{\beta}, \bm{\sigma_{\bm{i}}}, \sigma_{dur}]^T$. As the {\it Deterministic Model} considers a relation between consumption and production, normalization is needed to fix the scale. Thus, the consumption rate of weekdays $\lambda$ is fixed to 1 for identification. 

The following parameters are randomly distributed to account for individual heterogeneity: 
\begin{enumerate}
    \item {\it Value of time}. The value of time parameter $\rho_{1,n}$ has a unit of money/time (e.g., USD/hr). Consistent with the literature, it is assumed to follow a log-normal distribution (\cite{benakiva_bolduc_bradley}), given as:
    \begin{align}
        \rho_{1,n} = \exp(r_{\rho_1,n}),\ r_{\rho_1,n} \sim N(\mu_{\rho_1},\sigma_{\rho_1}^2) \label{eq:vot_distri}
    \end{align}

    \item {\it Value of inventory}. The value of inventory parameter $\rho_{3,n}$ has a unit of money/consumption-day (e.g., USD/consumption-day). It evaluates the monetary value of the psychological inventory. Define a variable $\kappa_n=\rho_{3,n} / \rho_{1,n}$, which has a unit of time/consumption-day. The new parameter $\kappa_n$ indicates that 1 consumption-day may be replenished in $\kappa_n$ unit of time. As $\lambda$ is fixed to 1, the following is required for feasibility: $\kappa_n \leq \min(FT_{WD,n}, FT_{WE,n})$. To satisfy this theoretical constraint, $\rho_{3,n}$ is given by:
    \begin{align}
        \rho_{3,n}=\rho_{1,n}\times \min{\left(FT_{WD,n}, FT_{WE,n}\right)} \times \frac{1}{1+\exp(r_{\kappa,n})},\ \ r_{\kappa,n}\sim N(\mu_\kappa,\sigma_\kappa^2) \label{eq:value_inventory_distri}
    \end{align}

    \item {\it Production function constant}. The production function constant $q_0$ is assumed to follow a normal distribution with parameters $\mu_{q_0}$ and $\sigma_{q_0}$: 
    \begin{align}
        q_{0,n}\sim N(\mu_{q_0},\sigma_{q_0}^2) \label{eq:q0_distri}
    \end{align}
\end{enumerate}

Therefore, the random parameters to be estimated are given by Equation (\ref{eq:random_dist}), with parameters $\bm{\mu_D}=[\mu_{\rho_1}, \mu_\kappa, \mu_{q_0}]$ and $\bm{\Omega_D}$, which is a diagonal matrix with $[\sigma_{\rho_1}^2, \sigma_\kappa^2, \sigma_{q_0}^2]$ on the diagonal:
\begin{align}
    \bm{\zeta_n}=\left[r_{\rho_1,n},r_{\kappa,n},q_{0,n}\right]^T\sim N(\bm{\mu_D},\mathbf{\Omega_D}) \label{eq:random_dist}
\end{align}

Finally, we show that $\rho_2 > \rho_3$ is required for boundedness and that the safety stock is 0 for all optimal solutions (Appendix \ref{appendixII:rho2bound}). Therefore, $\rho_2$ may be fixed to any arbitrary value greater than $\rho_3$. 

\subsection{Likelihood Function}\label{section:em_likelihood}
Individual likelihood is given by Equation (\ref{eq:indiv_likelihood}) as an integral over the distribution of the random parameters $\bm{\zeta_n}$, with a density function of $\Phi\left(\bm{\zeta_n}|\bm{\mu_D},\bm{\Omega_D}\right)$. Finally, the likelihood of observing patterns of all individuals in a sample ($\bm{Y}$) is a product of individual likelihood, as shown in Equation (\ref{eq:sample_likelihood}). 
\begin{align}
    &l\left(\bm{i_n},\bm{\delta_n},\bm{d_n}|\bm{\mu_D},\bm{\Omega_D},\bm{\xi},\bm{x_n}\right) =\int \left[ P(\bm{i_n},\bm{\delta_n},\bm{d_n}|\bm{\zeta_n},\bm{\xi},\bm{x_n})\times \Phi\left(\bm{\zeta_n}|\bm{\mu_D},\bm{\Omega_D}\right)\right]d\bm{\zeta_n} \label{eq:indiv_likelihood} \\
    &L\left(\bm{Y}\middle|\bm{\mu_D},\bm{\Omega_D},\bm{\xi},\bm{X}\right) =\prod_{n=1}^{N}l\left(\bm{i_n},\bm{\delta_n},\bm{d_n}|\bm{\mu_D},\bm{\Omega_D},\bm{\xi},\bm{x_n}\right) \label{eq:sample_likelihood}
\end{align}

\subsection{Estimation}\label{section:em_estimation}

To estimate the model, both maximum likelihood estimation and Hierarchical Bayesian estimation may be applied. No closed-form solution is available for the integral, and simulated maximum likelihood is computed. Furthermore, due to the large number of alternatives of the joint probability model of location and participation, a naive alternative sampling strategy is adopted \citep{GUEVARA2013185}.
\section{Model Computation}\label{section:computation}

The computation time with the Mosek solver to optimize the {\it Deterministic Model}, a MICP, still imposes a significant challenge on scalability \citep{mosek}. Preliminary testing on 100 individuals and 10 zones suggests that, with 128 sampled alternatives and 200 draws, a single evaluation of the likelihood function (Equation (\ref{eq:sample_likelihood})) takes on average 35 hours. 
Two computational enhancements are thus developed to address this issue, as introduced in the following section. The first enhancement approximates the non-linear production function with a piecewise linear function, thus transforming the model to a MILP. The second enhancement is the application of a feasibility and optimality condition-based algorithm to solve the MILP, as an alternative to the branch-and-bound algorithm applied by Gurobi \citep{gurobi}. 

\subsection{Linearization of Non-linear Functions}\label{piecewise_linear}

The production function in Equation (\ref{eq:dm_prod_def}) is approximated by a piecewise linear function. Note that in applying the {\it Deterministic Model} to compute probabilities for the {\it Empirical Model}, participation ($\bm{\delta}$) and location ($\bm{i}$) are fixed. Therefore, the production function can be rewritten as the following, where $C$ can be considered as a constant:
\begin{align}
    Q_t = Q(d_t, i_t)=C\cdot d_t^{q_1}, \quad\text{where} \ C=e^{q_0} A_{i_t,t}^{q_2} \nonumber
\end{align}
The piecewise linear approximation segments by the value of $d_t$. For instance, with three segments, the following form may be adopted, as shown by Equation (\ref{eq:piecewise_eq}). Since the piecewise linear function can be expressed as a combination of linear constraints, the MICP is now simplified to a MILP. 
\begin{align}
    Q_t\approx \begin{cases}
            {\widetilde{Q}}_t^1=C\cdot p_1 d_t\ \quad &0\leq d_t \leq d^1 \\ 
            {\widetilde{Q}}_t^2=C\cdot\left[p_2\left(d_t-d^1\right)+p_1 d^1\right]\ \quad &d^1\leq d_t \leq d^2 \\ 
            {\widetilde{Q}}_t^3=C\cdot\left[p_3\left(d_t-d^2\right)+p_2\left(d^2-d^1\right)+p_1 d^1\right]\ \quad &d^2 < d_t 
    \end{cases} \label{eq:piecewise_eq}
\end{align}

\begin{figure}[htpb]
    \centering
    \includegraphics[width=3 in]{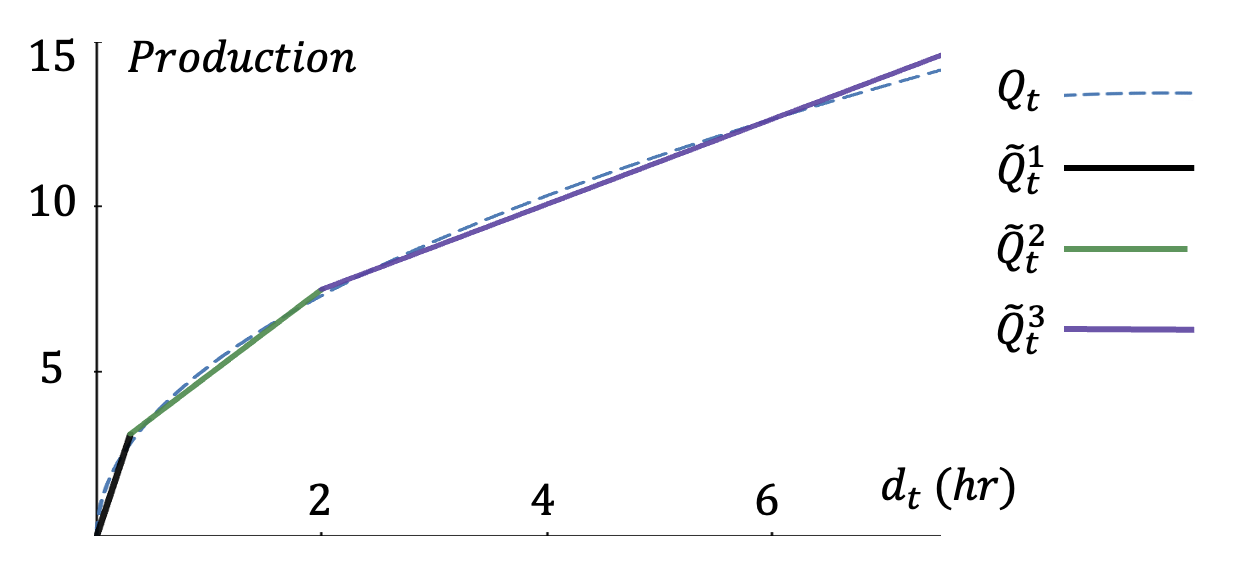}
    \caption{Piecewise Linear Approximation Illustration}
    \label{fig:piecewise_example}
\end{figure}

Figure \ref{fig:piecewise_example} illustrates one possible approximation for $q_0=-0.2$, $q_1=0.5$, $q_2=0.4$ and $A=100$. The approximation parameters are $p_1=2$, $p_2=0.5$, $p_3=0.25$, $d^1=0.3$ and $d^2=2$. The determination of the number of segments to apply is a modeling choice subject to consideration of both accuracy and computation time. Section~\ref{section:dm_tests} demonstrates experiment results for the linearization approximation. 

\subsection{Feasibility and Optimality Condition-based Algorithm}\label{section:alt_alg}

The MILP, although easier to solve than MICP, in general still cannot be solved using polynomial time algorithm, and thus requires significant computation time \citep{Bertsimas_Tsitsiklis_1997, JIMENEZCORDERO2022109570}. This section introduces an alternative algorithm to further improve the computational time of the standard branch-and-bound algorithm.

\subsubsection{Preliminaries}\label{section:alt_alg_preliminary}

As noted before, in the application of the needs-based model, {\it Deterministic Model} is used to optimize duration ($\bm{d}$), while the participation ($\bm{\delta}$) and location ($\bm{i}$) are fixed. Furthermore, the safety stock ($I_{min}$) is 0 for all optimal solutions, as shown in Appendix \ref{appendixII:rho2bound}. Consider solving the {\it Deterministic Model} with restrictions on $\bm{\delta}$ and $\bm{i}$. Let $k=\argmin_k I_{k}$. Lastly, consider the production function $Q_t=C p_1 d_t$ ($C=e^{q_0} A_{i_t,t}^{q_2}$). In what follows, we explain the approach for a linear production function but it can be generalized to piecewise linear production functions.

The slope of the objective function $V$ with respect to the decision variables $d_t \ \forall t\in\{1,\ldots,H\}$ can be derived as shown in Equation (\ref{eq:slope_eq}). The derivation may be found in Appendix \ref{appendixIII:obj_function_slope}:
\begin{align}
    \Delta_t = \frac{\partial V}{\partial d_t} = \begin{cases}
        \left(k-1+H-t\right)Cp_1\frac{\rho_3}{H}-\frac{\rho_1}{H}\ & t\geq\ k \\
        \left(k-1-t\right)Cp_1\frac{\rho_3}{H}-\frac{\rho_1}{H}\ & t< k
    \end{cases} \label{eq:slope_eq}
\end{align}

As a result, the following three observations can be made, which indicate that the slope of $d_t$ can be ordered. 
First, when both $t, t+1 \geq k$ or when both $t, t+1 < k$, we have $\Delta_t > \Delta_{t+1}$. Second, the slope of day $k$ is greater than all the others $t\neq k$: $\Delta_k > \Delta_t$. Thirdly, the slope of the last day is greater than the first day: $\Delta_H > \Delta_1$. In other words, anchoring at day $k$, the slope of the $d_t$'s decreases when the day goes forward ($t$ increases), but increases when the day goes backward ($t$ decreases). 
For example, for $H=7$ and $k=3$, the following order holds: $\Delta_3 > \Delta_4 > \Delta_5 > \Delta_6 > \Delta_7 > \Delta_1 > \Delta_2$. Thus, the ordering of the slope suggests a hierarchy of maximizing the activity duration by day, within the feasible region. As such, among the days when the activity may be performed (i.e. $t$ such that $\delta_t=1$), the maximization of the duration of day $k$ should be prioritized over the rest, followed by day $k + 1$, ..., $H$, $1$, ..., $k- 1$. 

\subsubsection{Algorithm}

Now an algorithm to solve the {\it Deterministic Model} with constrained $\bm{\delta}$ and $\bm{i}$ is summarized. For all $k$ such that $\delta_k=1$, apply the following steps: 
\begin{enumerate}
    \item Initialize the solution vectors $\bm{Q}=\bm{0}$ and $\bm{I} = \bm{0}$, all with length $H$.
    
    \item Set up an initial solution by computing the minimum production required for each day, which is the amount of inventory needed to support the consumption from the current day to the day before the next day of activity participation. Update $\bm{Q}$. Update $\bm{I}$ with $\bm{I}[k]=0$. Initial duration solution is given by $\bm{d}=\frac{\bm{Q}}{Cp_1}$. The current solution satisfies all constraints except the {\it Daily time constraint}. 

    \item Adjust duration day by day to achieve feasibility. For days where the {\it Daily time constraint} is violated, shift the excessive activity duration to prior days. If not doable, the problem is infeasible and the algorithm terminates. Otherwise, a feasible solution is constructed. 
    
    \item Adjust duration by the ordered slope $\Delta_i$ (Equation (\ref{eq:slope_eq})) to achieve optimality. For all days, check if a portion of the assigned duration may be shifted to days with a larger slope. For $t<k$, attempt to shift duration to day $t-1$, $t-2$, ..., 1, $H$, $H-1$, ..., $k$ where $\delta$ is constrained to be 1. For $t>k$, attempt to shift to day $t-1$, $t-2$, ..., $k$ where $\delta$ is constrained to be 1.
    
    \item Compute the inventory and optimal objective by solving Equation (\ref{eq:dm_obj}) to Equation (\ref{eq:dm_periodicity}) under the condition that $\bm{I}[k]=0$. Denote the current optimal objective as $V^*_k$.
\end{enumerate}

With the above, optimal solutions can be computed for all $k$ with $\delta_k=1$. The solution with the maximum $V^*_k$ will be the final optimal solution. Algorithm (\ref{alg:dm_solution}) presents the pseudo-code.

\begin{algorithm}
    \caption{Algorithm for Solving the Deterministic Model}\label{alg:dm_solution}
    \begin{algorithmic}
    \State $V^*, \bm{Q^*}, \bm{d^*}, \bm{I^*} \gets -\infty, None, None, None$
    \For{$k \gets 1$ to $H$ such that $\delta_k = 1$}
        \State $\bm{I}, \bm{Q} \gets \bm{0}, \bm{0}$ \Comment{Step 1}
        \State $\bm{a} \gets [t|\delta_t=1]$ \Comment{Step 2}
        \State $m \gets |\bm{a}|$
        \For{$i \gets 1$ to $m$}
            \State $\bm{Q}[\bm{a}\left[i\right]] \gets \sum_{t=\bm{a}\left[i\right]}^{\bm{a}\left[i+1\right]-1}\lambda_t \ \ \text{if} \ \ i < m \ \text{else} \ \sum_{t=\bm{a}\left[i\right]}^{H}\lambda_t+\sum_{t=1}^{\bm{a}\left[1\right]-1}\lambda_t$
        \EndFor
        \State $\bm{d} \gets \frac{\bm{Q}}{Cp_1}$
        \State $l \gets l$ where $\bm{a}[l] = k$ \Comment{Step 3}
        \For{$u \gets \{l-1, l-2,\ldots,1,m,m-1,\ldots,l+1\}$} 
            \If{$\left(\bm{d}\left[\bm{a}\left[u\right]\right]+TT_{i_{\bm{a}\left[u\right]}, \bm{a}\left[u\right]}>\bm{FT}[\bm{a}\left[u\right]]\right)$}
                \State $\bm{d}\left[\bm{a}\left[u-1\right]\right]+=\bm{d}\left[\bm{a}\left[u\right]\right]-\left(\bm{FT}\left[\bm{a}\left[u\right]\right]-TT_{i_{\bm{a}\left[u\right]}, \bm{a}\left[u\right]}\right)$
                \State $\bm{d}\left[\bm{a}\left[u\right]\right]=\bm{FT}\left[\bm{a}\left[u\right]\right]-TT_{i_{\bm{a}\left[u\right]}, \bm{a}\left[u\right]}$
            \EndIf
        \EndFor
        \If{$\bm{d}\left[k\right]+TT_{i_k, k}>\bm{FT}[k]$}
            \State Infeasible, terminate
        \EndIf
        \For{$u \gets \{l-1, l-2,\ldots,1,m,m-1,\ldots,l+1\}$}  \Comment{Step 4}
            \State $checks \gets \{u-1, u-2, \ldots, 1, m, m-1,\ldots,l+1,l\}$ if $u<l$ else $\{u-1,u-2,\ldots,l+1,l\}$
            \For{$j \gets checks$}
                \If{$\bm{FT}\left[\bm{a}[j]\right]-TT_{i_{\bm{a}[j]}, \bm{a}[j]}-\bm{d}\left[\bm{a}[j]\right]>0$}
                    \State $z=\min{\left(\bm{FT}\left[\bm{a}[j]\right]-TT_{i_{\bm{a}[j]}, \bm{a}[j]}-\bm{d}\left[\bm{a}[j]\right],\bm{d}\left[\bm{a}\left[u\right]\right]\right)}$
                    \State $\bm{d}\left[\bm{a}\left[u\right]\right]-=z$
                    \State $\bm{d}\left[\bm{a}[j]\right]+=z$
                \EndIf
            \EndFor
        \EndFor
        \State $\bm{Q} \gets Cp_1\bm{d}$
        \State $V^*_k, \bm{I} \gets $ Solve for Equation (\ref{eq:dm_obj}) to Equation (\ref{eq:dm_periodicity}) with $\bm{I}[k] = 0$ \Comment{Step 5}
        \If{$V^*_k > V^*$}
            \State $V^*, \bm{Q^*}, \bm{d^*}, \bm{I^*} \gets V^*_k, \bm{Q}, \bm{d}, \bm{I}$ 
        \EndIf
    \EndFor
    
    \end{algorithmic}
\end{algorithm}

To prove that Algorithm (\ref{alg:dm_solution}) finds the optimal solution, it suffices to prove that ($V^*_k, \bm{Q}, \bm{d}, \bm{I}$) after step 5 is optimal. By contradiction, ($V^*_k, \bm{Q}, \bm{d}, \bm{I}$) is not optimal. In other words, there exists an alternative solution ($\Tilde{V}_k, \Tilde{\bm{Q}}, \Tilde{\bm{d}}, \Tilde{\bm{I}}$) such that $\Tilde{V}_k > V^*_k$ and $\exists t$ where $d_t \neq \Tilde{d}_t$. Define $x_t = d_t - \Tilde{d}_t$. By Equation (\ref{eq:prod_depletion_same_total}) (Appendix \ref{appendixIII:obj_function_slope}):
\begin{align}
    \sum_{t=1}^{H}Q_t=\sum_{t=1}^{H}\lambda_t=\sum_{t=1}^{H}\Tilde{Q}_t \quad \Leftrightarrow \quad \sum_{t=1}^{H} C p_1 d_t =\sum_{t=1}^{H} C p_1 \Tilde{d}_t \quad \Leftrightarrow \quad \sum_{t=1}^H x_t = 0 \label{eq:alg_proof_sum_diff_equal_0}
\end{align}

With the shifting applied in the step 4 that adjusts duration toward $k$, the following inequalities hold. Otherwise, the alternative solution satisfies the maximizing priority better and would be the result of Algorithm (\ref{alg:dm_solution}) instead.
\begin{align}
    \forall h\in [k, H]: \sum_{t=k}^h x_t \geq 0 \label{eq:alg_proof_sum_x_t>=k} \\
    \forall h\in [1, k-1]: \sum_{t=h}^{k-1} x_t \leq 0 \label{eq:alg_proof_sum_x_t<k}
\end{align}

By the transformed objective function shown in Equation (\ref{eq:reformed_obj_fun}) (Appendix (\ref{appendixIII:obj_function_slope})) and the property of $I_{k}=0$ (proved in Appendix (\ref{appendixII:rho2bound})), the difference between $V^*_k$ and $\Tilde{V}_k$ is given as:
\begin{align}
     V^*_k - \Tilde{V}_k = &\frac{\rho_3}{H}\left[ HI_k+\sum_{t=1}^{H}\frac{\lambda_t}{2}+\sum_{t=k}^{H}{\left(k-1+H-t\right)(C p_1 d_t-\lambda_t)}+\sum_{t=1}^{k-1}\left(k-1-t\right)\left(C p_1 d_t-\lambda_t\right) \right] \nonumber \\
        & -\left[\frac{\rho_1}{H}\sum_{t=1}^{H}\left(d_t+\delta_tTT_{i_t,t}\right)+\rho_2I_k+\frac{1}{H}\sum_{t=1}^{H}\left(\delta_t{TC}_{i_t,t}\right)\right] \nonumber \\
        & -\frac{\rho_3}{H}\left[ HI_k+\sum_{t=1}^{H}\frac{\lambda_t}{2}+\sum_{t=k}^{H}{\left(k-1+H-t\right)(C p_1 \Tilde{d}_t-\lambda_t)}+\sum_{t=1}^{k-1}\left(k-1-t\right)\left(C p_1 \Tilde{d}_t-\lambda_t\right) \right] \nonumber \\
        & +\left[\frac{\rho_1}{H}\sum_{t=1}^{H}\left(\Tilde{d}_t+\delta_tTT_{i_t,t}\right)+\rho_2I_k+\frac{1}{H}\sum_{t=1}^{H}\left(\delta_t{TC}_{i_t,t}\right)\right] \nonumber \\ 
        = &\frac{\rho_3}{H}\left[\sum_{t=k}^{H}{\left(k-1+H-t\right)C p_1 x_t}+\sum_{t=1}^{k-1}\left(k-1-t\right)C p_1 x_t \right] -\frac{\rho_1}{H}\sum_{t=1}^{H}x_t  
\end{align}

The equation can be further simplied by the property shown in Equation (\ref{eq:alg_proof_sum_diff_equal_0}): 
\begin{align}
    V^*_k - \Tilde{V}_k =\frac{\rho_3}{H}C p_1 \left[\sum_{t=k}^{H}\left(H-t\right) x_t - \sum_{t=1}^{k-1} t x_t \right] \label{eq:alg_proof_diff_in_V_with_summation}
\end{align}

Equation (\ref{eq:alg_proof_sum_x_t>=k}) suggests that the following is true: 
\begin{align}
    & \sum_{t=k}^k x_t + \sum_{t=k}^{k+1} x_t + ... + \sum_{t=k}^{H-1} x_t \geq 0 \nonumber \\
    \Leftrightarrow \quad & (H-k) x_k + (H-k-1)x_{k+1} + ... + x_{H-1} \geq 0 \nonumber \\ 
    \Leftrightarrow \quad & \sum_{t=k}^H (H-t) x_t \geq 0 
\end{align}

Similarly, Equation (\ref{eq:alg_proof_sum_x_t<k}) implies the following:
\begin{align}
    \sum_{t=1}^{k-1} t x_t \leq 0 
\end{align}

Therefore, the following holds for Equation (\ref{eq:alg_proof_diff_in_V_with_summation}):
\begin{align}
    V^*_k - \Tilde{V}_k \geq 0
\end{align}

Therefore, the alternative solution ($\Tilde{V}_k, \Tilde{\bm{Q}}, \Tilde{\bm{d}}, \Tilde{\bm{I}}$) is not optimal. Hence, the solution ($V^*_k, \bm{Q}, \bm{d}, \bm{I}$) computed by Algorithm (\ref{alg:dm_solution}) is the optimal solution for $\bm{I}[k]=0$.

\section{Numerical Experiments}\label{section:numerical_experiment}

This section describes numerical experiments to examine performance of the proposed model.
The computational performance of the {\it Deterministic Model} and enhancements are first examined in Section \ref{section:dm_tests}. 
Following this, Monte Carlo experiments are performed on the overall needs-based model in Section \ref{section:monte_carlo_exp} to examine the concavity of the log-likelihood function and the recovery of true parameters. 
Finally, Section \ref{section:e_commerce_demo} demonstrates the potential to model e-commerce activity.

\subsection{Deterministic Model Test}\label{section:dm_tests}

This section presents the results of the two computational enhancements presented in Section \ref{section:computation}. The model is implemented in Julia \citep{Julia-2017}. Mosek and Gurobi solvers are used respectively for the MICP and MILP \citep{gurobi, mosek}. 

The first enhancement, linearization of the production function, is tested. Piecewise linear approximations with 1 up to 7 segments are applied, and each segment is tested with the parameters shown in Table \ref{table:boosterI_param}. Parameters $\gamma$, $q_0$, and $q_2$, which are related to the rate of consumption and production, are varied to ensure the generalizability of the results. Therefore, for each segment, 100 sets of tests with different sets of parameters are conducted. For each segment test, to find the piecewise linear function parameters to approximate the Cobb-Douglas function, a naive approach is taken: (1) generate 800 equidistance points between 0.01 and 8.0; (2) evaluate both the Cobb-Douglas and the piecewise linear functions at the 800 points, and denote the values respectively as $Q$ and $\hat{Q}$; (3) find the piecewise linear parameters that minimize $\sum_{k=1}^{800} (Q_k - \hat{Q}_k)^2$. Finally, the independent variables are fixed to specific values. Free time is set as 2 hr for weekdays and 6 hr for weekend days. Furthermore, for testing purposes, it is assumed that all locations have an attractiveness of 100, and travel time/cost are the same across all days, respectively 0.5 hr and 5 USD one-way. 
\begin{table}[ht]
\center
\TABLE
{Computation Enhancement I Test Parameters \label{table:boosterI_param}}
{\begin{tabular}{ |c|c|c|c| } 
\hline
Symbol & Name & Value & Varied (Y) \\
\hline
$\lambda$  	& weekday consumption rate       & 1 consumption-day           & \\ 
$\gamma$ & weekend/weekday consumption rate ratio  & $\{0.6, 0.8, 1.0, 1.2, 1.4\}$  & Y \\ 
$\rho_1$ 	& value of time         & 30 \$/hr                          &  \\
$\rho_3$  	& value of inventory     & 15 \$/consumption-day              &  \\
$\rho_2$  	& value of safety stock		  & $2\rho_3$ (for boundedness)   &  \\
$q_0$       & production function constant  & $\{-0.4, -0.2, 0, 0.2, 0.4\}$ & Y \\ 
$q_1$     & elasticity of duration     & 0.5        & \\
$q_2$ 	&elasticity of location attractiveness  	&$\{0.2, 0.4, 0.6, 0.8\}$ & Y \\ 
\hline
\end{tabular}}
{}
\end{table}

\begin{figure}[htpb]
    \centering
    \includegraphics[width=6.5 in]{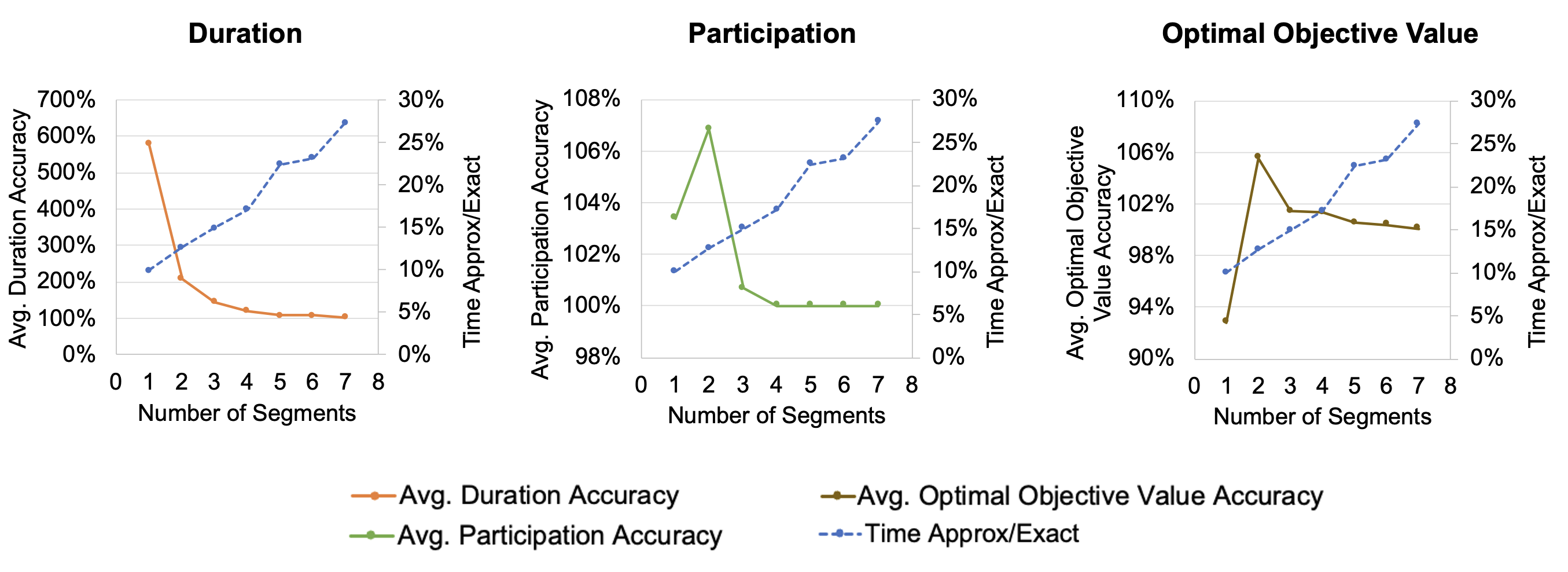}
    \caption{Piecewise Linear Approximation Test}
    \label{fig:piecewise_test}
\end{figure}

Two metrics are compared: accuracy and time. Accuracy compares the average duration, average weekly participation, and optimal objective value computed with the piecewise linear function against those computed with the Cobb-Douglas (CD) function. It is defined as $a=\frac{Piecewise(k)}{CD(k)}$, where $k$ stands for the measure of comparison. A value of 100\% suggests an exact approximation to the Cobb-Douglas function. On the other hand, the comparison of time is based on an average of 50 runs: $b=\frac{\frac{1}{50} \sum_{l=1}^{50} T^l_{Piecewise}}{\frac{1}{50} \sum_{l=1}^{50} T^l_{CD}}$. For each of the 100 tests for each segment, both $a$ and $b$ are computed. Figure \ref{fig:piecewise_test} presents the results, where each point represents the average of the 100 tests for a particular number of segments.

The results demonstrate that the speedup is around 70 - 90\% from the Cobb-Douglas function. The accuracy of duration, participation, and optimal objective value can be well maintained with three segments, which yields a speedup of around 85\%. These results indicate that the Cobb-Douglas function may be replaced by a piecewise linear approximation for its benefits in computational speed. 

Next, the feasibility and optimality condition-based algorithm is tested, using the same set of parameters from Table \ref{table:boosterI_param} but with several changes. First, the linear production function is used, consistent with the development in Section \ref{section:alt_alg}. The $p_1$ parameter is assigned a value of 0.5. Second, as noted in Section \ref{section:alt_alg}, $\bm{\delta}$ is treated as a constrained input. For each one of the 100 tests with different parameters, 25\% of all possible $\bm{\delta}$ are tested. Thus, $100\times 25\% \times 2^7 = 3200$ tests are conducted to compare the computation time of Gurobi and the algorithm presented in Section \ref{section:alt_alg}. As a result, the average computation time from Gurobi is 0.00596 seconds, while the algorithm proposed has an average time of 0.0000376 seconds. Therefore, we obtain around 99\% speedup from Gurobi, and hereby achieve scalability.  

It should be noted, however, the form of the production function is a modeling choice and relies on empirical data to validate. While we propose the Cobb-Douglas function for its wide applicability in economic research and developed the piecewise linear approximation for computation efficiency, further investigations need to be conducted on the proper form to apply. In some cases, a simple linear function may suffice.

\subsection{Monte Carlo Experiment}\label{section:monte_carlo_exp}

The proposed needs-based model is tested on synthetic data generated using Monte Carlo simulation. A sample of 1500 individuals and 10 zones is first generated as described below.

Independent variables are generated. First, two size measures that affect the location choice (Equation (\ref{eq:size_measure})) are considered: retail employment with a uniform distribution between 50 and 100, and the zonal area with a uniform distribution between 0.1 and 2 (mile$^2$). Retail employment density is used as a measure of location attractiveness. The travel time matrix is generated following three steps: (1) draw random numbers from a uniform distribution between $5/60$ and $1$ hr; (2) adjust the matrix to achieve symmetry by setting $TT[j,i]=TT[i,j] \ \forall j>i$; (3) perturb all elements in the matrix by multiplying each value to a random number uniformly distributed between 0.9 and 1.1. Travel cost is generated based on travel time as $TT_{i,j}\times u_{i,j}\times 12.8\$/hr$, where $u_{i,j}$ is a perturbation factor uniformly distributed between 0.9 and 1.1. The value of $12.8\$/hr$ is adopted from \cite{aaa_2021} and assumes a driving speed of 20 mph. The free time of individuals as an independent variable is also generated. Free time of weekday follows a distribution of $\frac{8\ hr\ }{1+\exp(r_{FT,WD,n})},r_{FT,WD,n}\sim\ N\left({1.0,\ 0.5}^2\right)$, which is guaranteed to be smaller than 8 hr, assuming a working individual. Similarly, the free time of weekend days is generated as $\frac{16\ hr}{1+\exp(r_{FT,WE,n})},r_{FT,WE,n}\sim\ N\left(0.8,\ {0.4}^2\right)$ to guarantee that the value is smaller than 16 hr. Each individual is randomly assigned a home location. 

Parameter values are assigned or randomly generated as summarized by Table \ref{table:mc_test_param}. A linear production function is used. The variance of the error component $\eta_{\bm{i}}$, namely $\sigma_{\bm{i}}$, is assumed to be the same for all location patterns. Furthermore, it is assumed that all location patterns considered only consist of one location to further constrain the computation time. In other words, the same location will be visited for all activity participation over the planning horizon. Thus, the choice set of location patterns is significantly reduced. 

\begin{table}[ht]
\center
\TABLE
{Monte Carlo Experiment Parameters \label{table:mc_test_param}}
{\begin{tabular}{ |c|c|c| } 
\hline
Symbol & Name & Value \\
\hline
$\lambda $  	& weekday consumption rate  & 1  \\ 
$\gamma$        & weekend/weekday consumption rate ratio & 1.2 \\ 
$p_1$           & elasticity of activity duration (linear)  & 0.8 \\ 
$q_2$           & elasticity of location attractiveness     & 0.5 \\
$\mu$           & scale parameter of location and participation choice            & 0.2 \\ 
$\beta_{RE}$    & retail employment size measure parameter of location and participation choice  & 0.5 \\ 
$\beta_{Area}$    & zone area size measure parameter of location and participation choice & 1.0 \\ 
$\sigma_{\bm{i}}$    & variance of error component of location and participation 
 choice    & 5.0 \\ 
$\sigma_{dur}$      & variance of the multiplicative error of duration probability         & 0.2 \\
$r_{\rho_1,n}$    & variable transformed to value of time (Equation (\ref{eq:vot_distri}))    & $\sim\ N\left(3.0,\ {1.0}^2\right)$ \\
$r_{\kappa,n}$    & variable transformed to value of inventory (Equation (\ref{eq:value_inventory_distri}))   & $\sim\ N({1.0,0.5}^2)$ \\
$q_{0,n}$           & production function constant  & $\sim\ N(-0.5,\ {0.5}^2)$ \\
\hline 
\end{tabular}}
{}
\end{table}

\begin{figure}[ht]
    \begin{subfigure}{.5\textwidth}
        \centering
        \includegraphics[width=.75\linewidth]{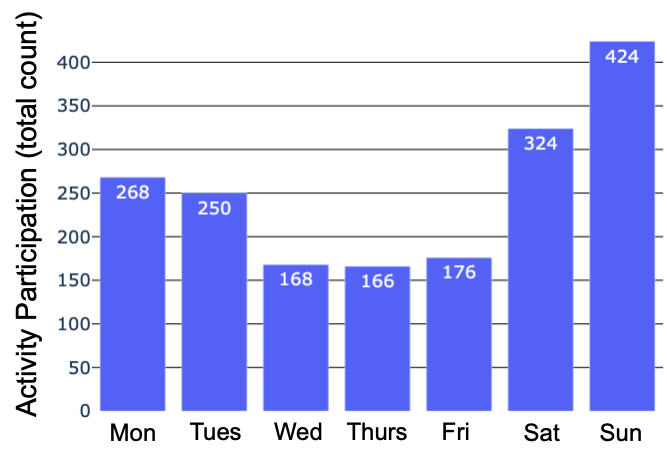}  
        \caption{Total participation} \label{fig:synthesized_total_participation}
    \end{subfigure}
    \begin{subfigure}{.5\textwidth}
        \centering
        \includegraphics[width=.75\linewidth]{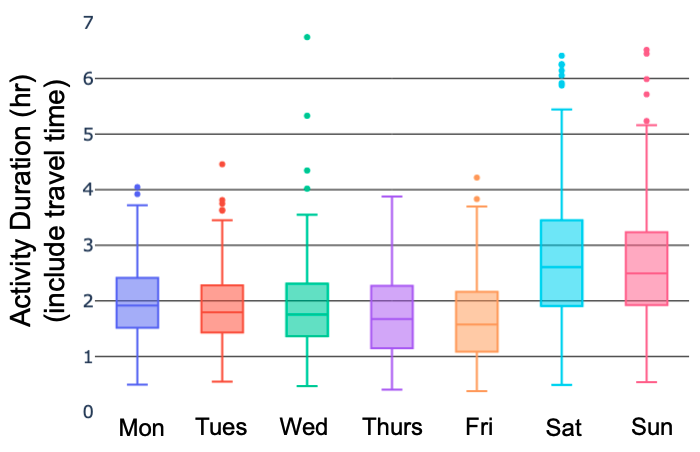}  
        \caption{Activity duration distribution} \label{fig:synthesized_avg_duration}
    \end{subfigure}
    \newline
    \begin{subfigure}{0.475\textwidth}
        \centering
        \includegraphics[width=.8\linewidth]{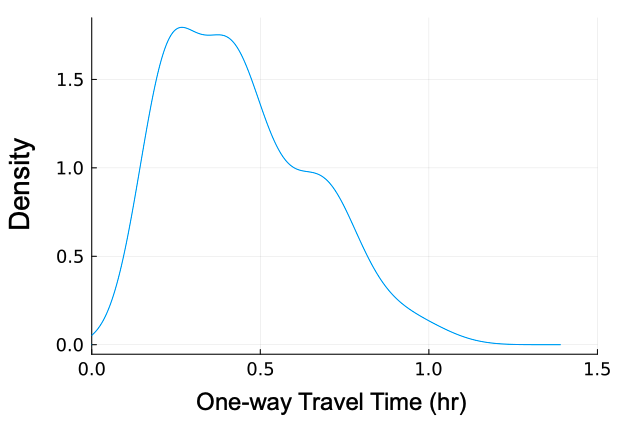}
        \caption{One-way travel time distribution} \label{fig:synthesized_tt}
    \end{subfigure}
    \begin{subfigure}{0.475\textwidth}
        \centering
        \includegraphics[width=.8\linewidth]{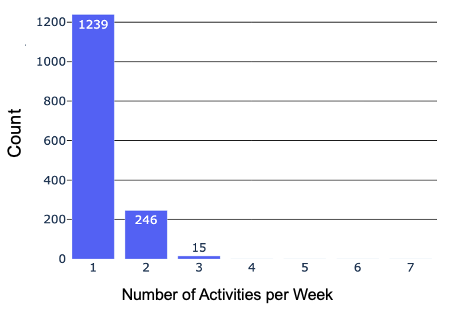}
        \caption{Activity participation per week} \label{fig:synthesized_num_participation_per_week}
    \end{subfigure}
    \caption{Pattern of Synthesized Sample for Monte Carlo Experiment} \label{fig:synthesized_pattern}
\end{figure}

Using the needs-based model, activity patterns are generated for the 1500 individuals. Figure~\ref{fig:synthesized_pattern} shows the characteristics of the activity patterns synthesized. Although parameters are not calibrated against real-world data, the generated activity patterns may be considered reasonably similar to grocery shopping activity. Activities are conducted more during the weekend (Figure~\ref{fig:synthesized_total_participation}), especially on Sunday, which is consistent with the report by \cite{klee_2020} on grocery shopping. According to the American Time Use Survey, the average time spent on activities of purchasing goods and services is 1.65 hr for weekdays and 1.81 hr for weekends and holidays, including travel time \citep{us_bureau_Statistics_2023}. The synthesized activities have an average duration within a similar range and show an increase in weekend duration as well (Figure~\ref{fig:synthesized_avg_duration}). The average one-way travel time to grocery stores in the U.S. in 2012 is reported to be 15 minutes \citep{Hamrick_Hopkins_2012}. Figure~\ref{fig:synthesized_tt} shows the distribution of the one-way travel time of the synthesized activities, which has a mean of 26.5 minutes. Finally, the average weekly grocery shopping trips per household is reported to be around 1.5 to 1.7 (2019 - 2022) \citep{Ozbun_2023}. The average weekly trip among the synthesized population is 1.18, with a distribution shown in Figure~\ref{fig:synthesized_num_participation_per_week}. 
Note that this is only intended to be a proof-of-concept to demonstrate reasonableness of the model. A next step would be estimation and validation using empirical data of multi-day travel diaries, which we defer to future research.

\begin{figure}[htpb]
    \centering
    \begin{subfigure}[b]{0.8\textwidth}
        \centering
        \includegraphics[width=4.6 in]{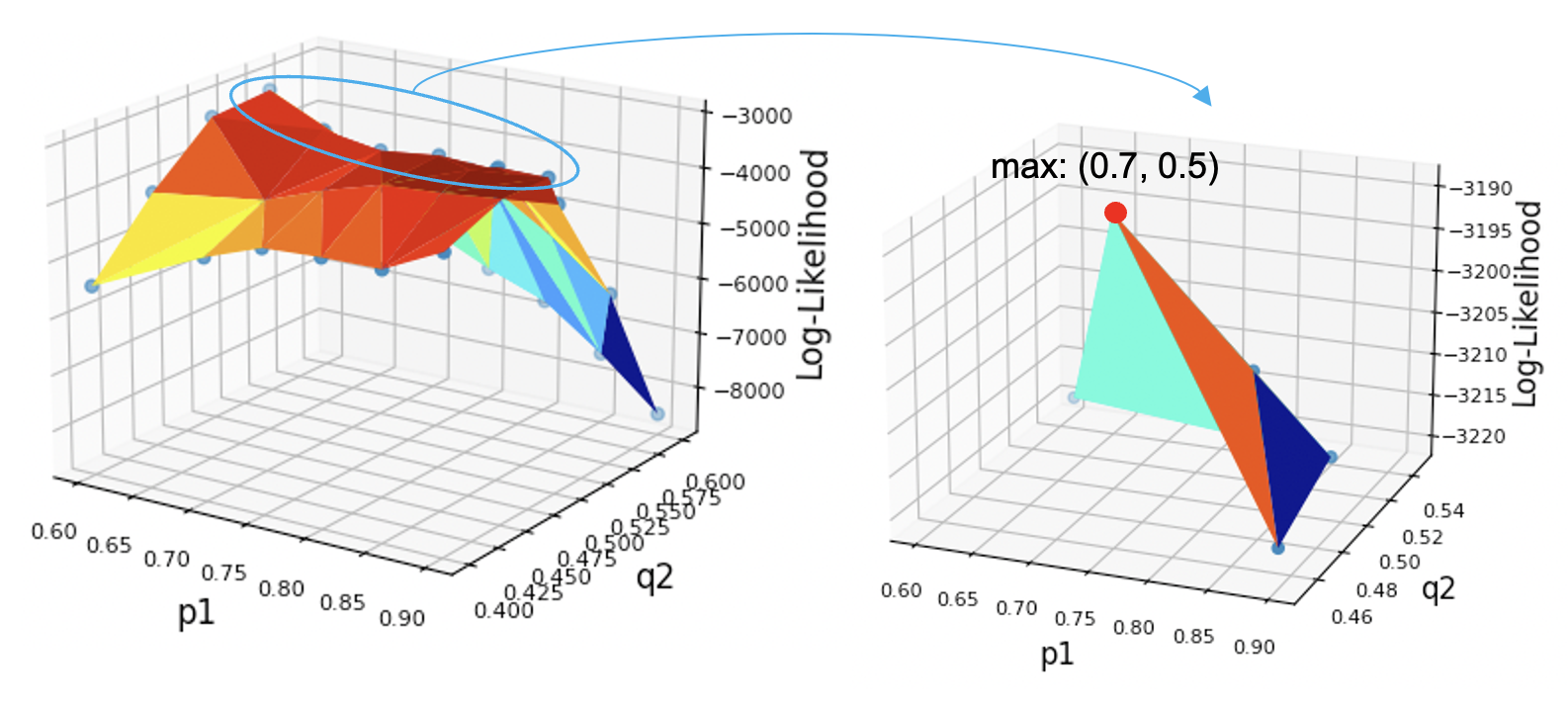}        
        \caption{1000 individuals, 500 draws} \label{fig:ll_shape_1000_500}
    \end{subfigure}

    \begin{subfigure}{0.8\textwidth}
        \centering
        \includegraphics[width=4.6 in]{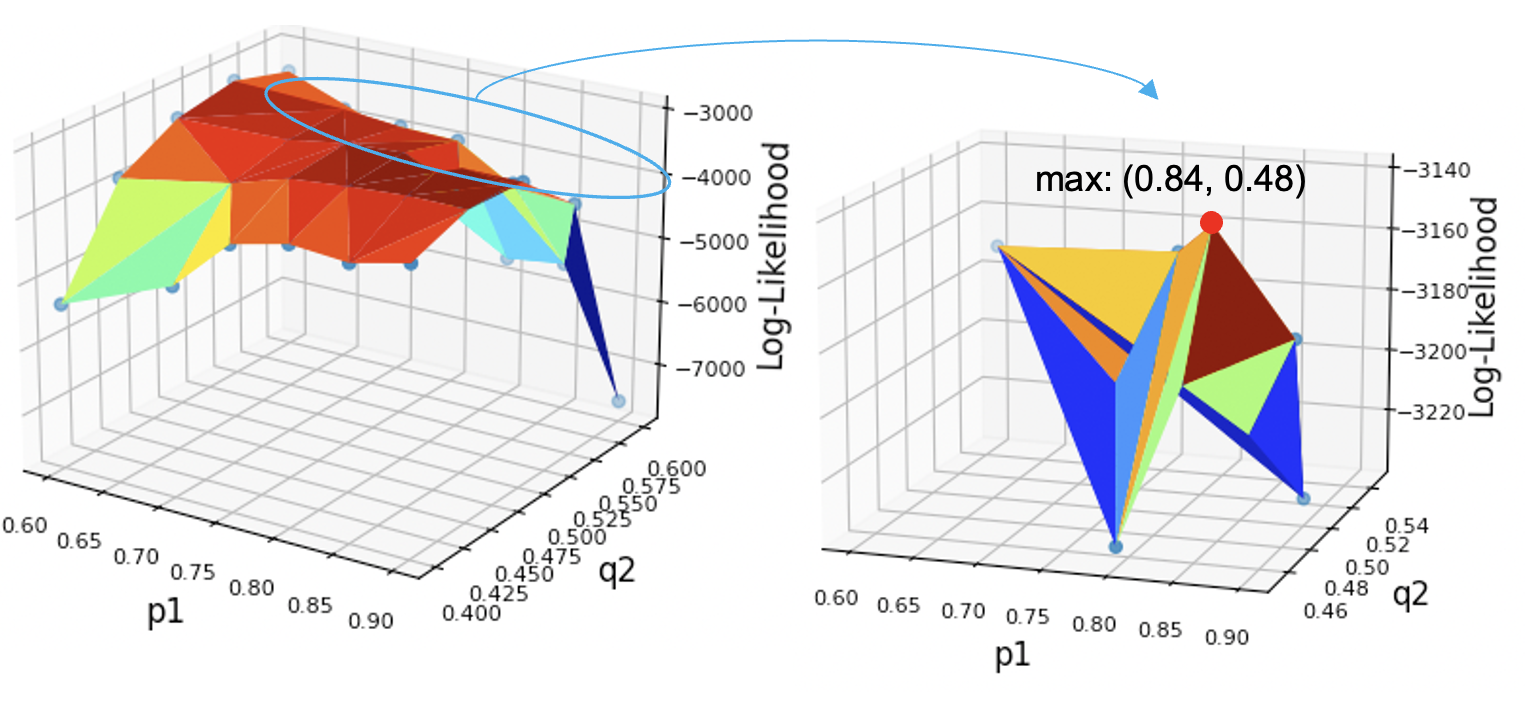}
        \caption{1000 individuals, 1000 draws} \label{fig:ll_shape_1000_1000}
    \end{subfigure}

    \begin{subfigure}{0.8\textwidth}
        \centering
        \includegraphics[width=4.6 in]{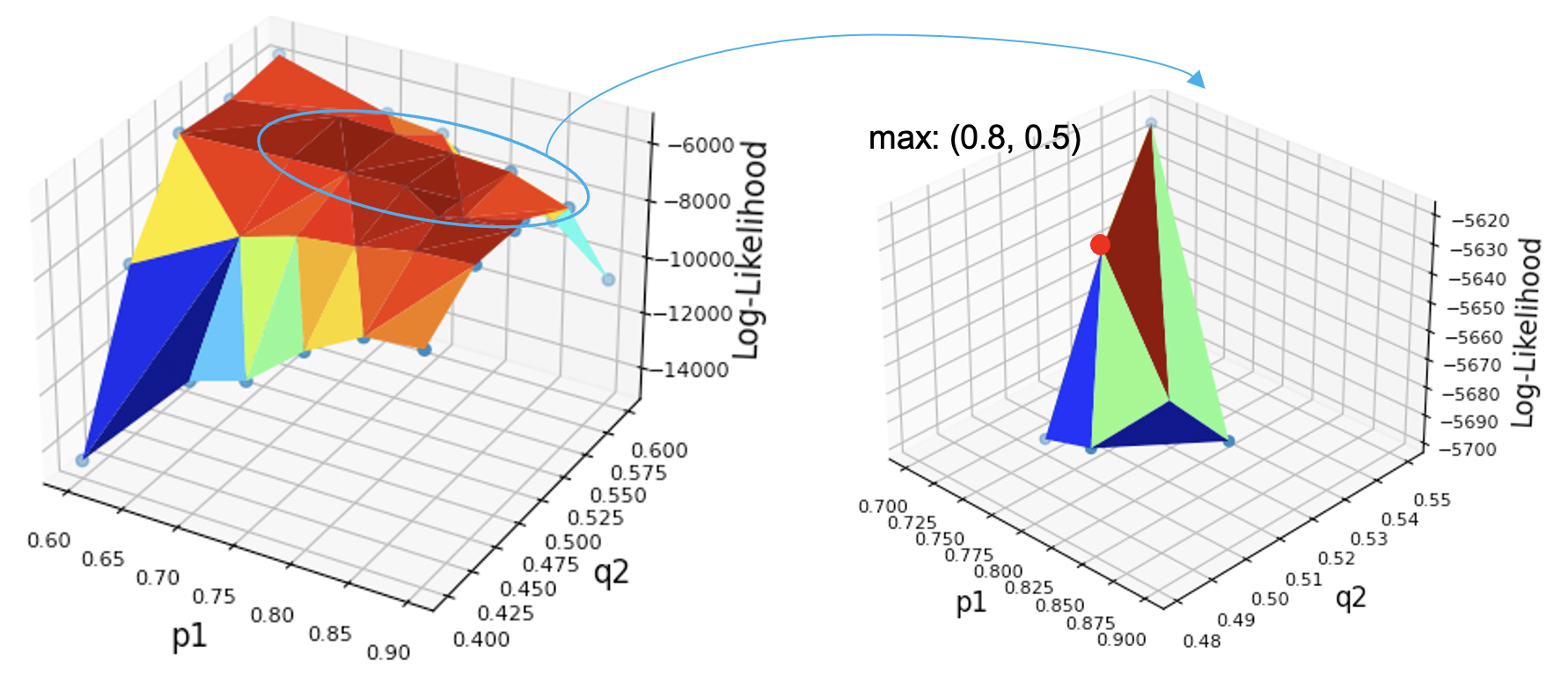}
        \caption{1500 individuals, 500 draws} \label{fig:ll_shape_1500_500}
    \end{subfigure}
    \caption{Shape of Log-likelihood with respect to $p_1$ and $q_2$, under the Impacts of Number of Samples and Number of Draws (red dots are points where the log-likelihood is the maximum)}
    \label{fig:ll_shape}
\end{figure}

Two tests are performed: the examination of the concavity of the log-likelihood function and the recovery of the true parameters. First, the concavity of the log-likelihood function is examined, focusing on the impacts of two factors: the number of samples and the number of draws for computing the simulated log-likelihood. The shape is plotted against changes in two parameters: $p_1$ and $q_2$, whose true values are respectively 0.8 and 0.5. Three figures are created with different numbers of samples and draws, as shown in Figure \ref{fig:ll_shape}. The result of 1000 individuals and 500 draws (Figure \ref{fig:ll_shape_1000_500}) does not show a concave shape and two peaks are shown. The maximum occurs at $p_1=0.7$ and $q_2=0.5$. When the number of draws is increased from 500 to 1000, the shape of the log-likelihood function is concave but flat with a maximum at $p_1=0.84$ and $q_2=0.48$ (Figure \ref{fig:ll_shape_1000_1000}). When more samples are included, on the other hand, the global maximum is achieved at the true value (Figure \ref{fig:ll_shape_1500_500}). The results indicate that, while increasing the number of draws improves the concavity in shape, to achieve a global maximum nearby the true value, a sufficient number of samples is critical. 

\begin{figure}[htpb]
    \centering
    \includegraphics[width=3 in]{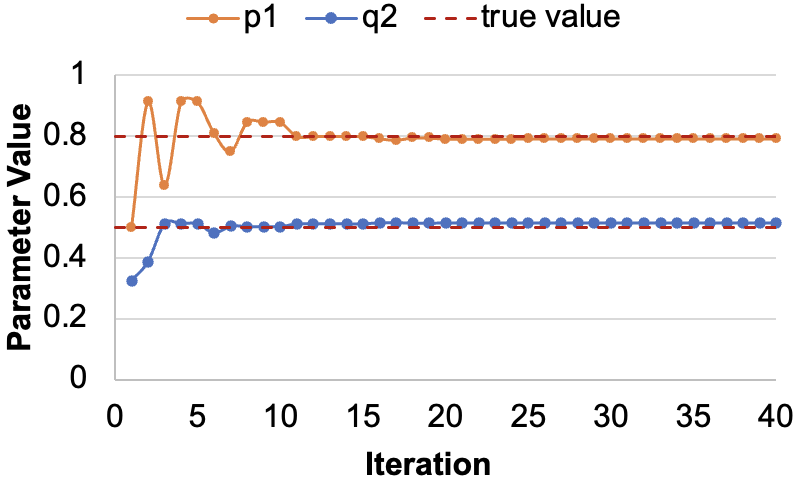}
    \caption{Maximum Log-likelihood Test: Changes of Parameter Estimate Over Iterations}
    \label{fig:mle_param_change}
\end{figure}

In addition, a test of the recovery of true parameters is performed with 1500 individuals for a sufficient sample and 1000 draws for stable results. The Julia Optim package is used for the log-likelihood maximization, with a maximum of 40 iterations configured \citep{mogensen2018optim}. Except for the testing parameters $p_1$ and $q_2$, all other parameters are fixed to their true value. Figure \ref{fig:mle_param_change} shows the changes in the estimates of the two parameters. Around iteration 10, the estimates have approached the true value. After 40 iterations, the estimates are 0.791 and 0.515 respectively for $p_1$ and $q_2$, which are close to the true value of 0.8 and 0.5. This test is performed on a Dell with Intel Xeon Gold 6226R processor, 448 GB memory, and 32 cores (64 CPU). The computation time of 40 iterations is 2 days 1.5 hours. Parallelization is implemented to reduce the computation time. 

\subsection{E-Commerce Demonstration}\label{section:e_commerce_demo}

\begin{figure}[ht]
    \begin{subfigure}{0.475\textwidth}
        \centering
        \includegraphics[width=.9\linewidth]{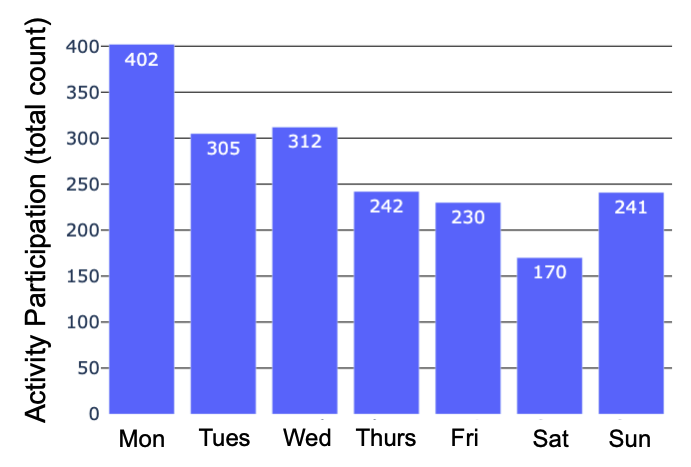}
        \caption{Total participation} \label{fig:synthesized_ecommerce_part}
    \end{subfigure}
    \begin{subfigure}{0.475\textwidth}
        \centering
        \includegraphics[width=.9\linewidth]{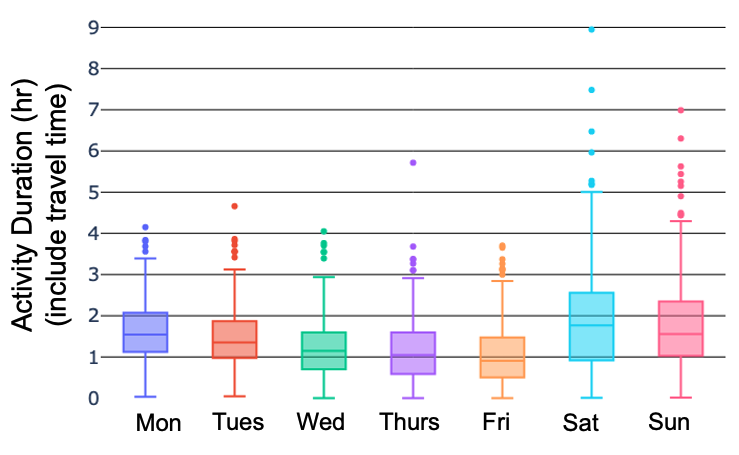}
        \caption{Activity duration distribution} \label{fig:synthesized_ecommerce_duration}
    \end{subfigure}
    \caption{Pattern of Synthesized Sample for E-Commerce Demonstration}\label{fig:ecommerce_demonstration}
\end{figure}

Finally, to demonstrate the potential of the proposed model, the inputs and parameters used in Section \ref{section:monte_carlo_exp} are updated to demonstrate possible e-commerce activity patterns. First, only one location, namely online, is considered as the candidate location. Second, no measure of size or attractiveness is proposed for online shopping to the best of the authors' knowledge, and therefore, instead of employment density, the attractiveness is arbitrarily fixed to 100. Third, online shopping requires no travel, and therefore travel time and cost are set to 0. The free time of both weekday and weekend days followed the same distribution described in Section \ref{section:monte_carlo_exp}. Size measures are not used for Equation (\ref{eq:loc_part_sys_util}) in this exercise. In terms of the individual parameters, $\gamma$ is fixed to 1.4, assuming a larger depletion rate over the weekend. The scale parameter $\mu$ of the joint location and participation choice is decreased to 0.1, assuming a less determinant decision. The rest of the parameters have the same value or distribution as summarized in Table \ref{table:mc_test_param}. Figure \ref{fig:ecommerce_demonstration} shows the daily participation and distribution of activity durations by day of the week, among the 1500 individuals. Thus, Monday has the highest e-commerce activity participation, and in general, the patterns appear reasonable in the context of e-commerce activity \citep{statista_2017}. 

\section{Conclusion}\label{section:conclusion}

This paper proposed a multi-day needs-based model of travel demand. Psychological needs are considered as the underlying drivers of activities. An analytical model was developed to model the decisions on participation, location, and activity duration. The level of need satisfaction is modeled as an inventory that depletes over time but can be recharged by activities. The proposed needs-based model consists of a {\it Deterministic Model} and an {\it Empirical Model}. The {\it Deterministic Model} is an optimization model, whose decision variables are activity participation, location, and duration, and maximizes the utility of the psychological inventory. The {\it Empirical Model} further captures the heterogeneity and stochasticity and proposes the probabilistic models, the likelihood function, the inputs, the parameter distributions, and estimation methods for application. Multiple computational enhancements were proposed to ensure scalability of the model and numerical experiments were performed to test the efficacy of these enhancements. Monte Carlo experiments were also conducted to examine the concavity of the log-likelihood function and the recovery of selected model parameters via a maximum likelihood estimator.

There are several avenues for future research. First, although grounded in psychological theory, the model needs to be estimated and validated on empirical data consisting of multi-day activity diaries. Second, for the Monte Carlo experiment, the simplification that the same location is to be visited for all participation needs to be further relaxed. While this simplification may not be too unrealistic for certain types of activities (e.g. shopping), it may not be always appropriate. Instead, other means to control the computation time may be developed: e.g., a higher-level location choice set generation model to limit the choice set size. Third, the Monte Carlo experiments need to be conducted on the full set of parameters. Identification issues should be examined. The tests to recover the full set of parameters need to be conducted. Finally, the activity and need were assumed to have a one-to-one relationship. However, many-to-many relationships exist: a need may be satisfied by multiple activities; conversely, an activity may impact multiple needs.

\ACKNOWLEDGMENT{%
    This research was funded by the Shenzhen Urban Transport Planning Center CO.,LTD. The authors would like to thank their generous support. 
}%

\begin{APPENDICES}
    \section{Boundedness of Deterministic Model}\label{appendixII:rho2bound}

\subsection{Theoretical Bound of Value of Safety Stock Parameter}\label{appendixII_1:rho2_boundproof}
It is necessary to ensure that the optimization problem given by Equation (\ref{eq:dm_obj}) to Equation (\ref{eq:dm_i_value}) is bounded. Consider an arbitrary feasible and bounded solution: $\left(\left\{\delta_t\right\}_{t=1}^H, \left\{d_t\right\}_{t=1}^H, \left\{i_t\right\}_{t=1}^H, \left\{I_t\right\}_{t=1}^H \right)$. Consider an arbitrary constant $c > 0$. Then $\left(\left\{\delta_t\right\}_{t=1}^H, \left\{d_t\right\}_{t=1}^H, \left\{i_t\right\}_{t=1}^H, \left\{I_t + c\right\}_{t=1}^H \right)$ is also a feasible solution. For the new solution to be bounded, the following condition needs to be satisfied: 
\begin{align}
    &\frac{\rho_3}{H}\ \sum_{t=1}^H \left[I_t + c +Q(d_t, i_t) -\frac{\lambda_t}{2}\right] - \left[\frac{\rho_1}{H}\ \sum_{t=1}^H (d_t + \delta_t TT_{i_t,t}) + \rho_2 \min_{t=1,\ldots,\ H}{(I_t+c)} + \frac{1}{H}\ \sum_{t=1}^H (\delta_t TC_{i_t, t}) \right] \nonumber\\
    =&\frac{\rho_3}{H}\sum_{t=1}^{H}\left[I_t+Q_t-\frac{\lambda_t}{2}\right]+\frac{\rho_3}{H}\sum_{t=1}^{H}c-\frac{\rho_1}{H}\sum_{t=1}^{H}\left(d_t+\delta_tTT_{i_t,t}\right)-\rho_2 \min_{t=1,\ldots,\ H}{(I_t+c)}-\frac{1}{H}\sum_{t=1}^{H}\left(\delta_t{TC}_{i_t,t}\right) \nonumber\\
     < & + \infty 
\end{align}

Rewrite the above constraint, the following needs to be satisfied: 
\begin{align}
    \frac{\rho_3}{H}\sum_{t=1}^{H}\left[I_t+Q_t-\frac{\lambda_t}{2}\right]-\frac{\rho_1}{H}\sum_{t=1}^{H}\left(d_t+\delta_tTT_{i_t,t}\right)-\rho_2\min_{t=1,\ldots,\ H}I_t-\frac{1}{H}\sum_{t=1}^{H}\left(\delta_t{TC}_{i_t,t}\right)+c(\rho_3-\rho_2)<+\infty  
\end{align}

For $\rho_2 < \rho_3$, $c$ could be arbitrarily large and results in unboundednes. Therefore, the $\rho_2$ parameter need to have the following theoretical bound: 
\begin{align}
    \rho_2 > \rho_3 \label{eq:rho2_rho3_constraint}
\end{align}

\subsection{Proof of Safety Stock Value}
Furthermore, when the model is bounded, the value of safety stock will always be zero: $I_{min}=0$. The following section presents the proof. Suppose $k=\argmin_{i=1,\ldots,H}\ I_i$ ($I_k = I_{min}$). First, rewrite the benefit expression of the objective function as a function of $I_k$: 
\begin{align}
    \frac{\rho_3}{H}\sum_{t=1}^{H}\left[I_t+Q_t-\frac{\lambda_t}{2}\right] = \frac{\rho_3}{H}\left[\left(I_k+Q_k-\frac{\lambda_k}{2}\right) + \sum_{t=1}^{k-1}\left(I_t+Q_t-\frac{\lambda_t}{2}\right) + \sum_{t=k+1}^{H}\left(I_t+Q_t-\frac{\lambda_t}{2}\right)\right] \label{eq:benefit_rewrite}
\end{align}

Conditioned on the value of $t$, $I_t$ can be expressed as a function of $I_k$, :
\begin{align}
    &I_t =I_k+\sum_{i=k}^{t-1}\left(Q_i-\lambda_i\right)  &\forall t \geq k \label{eq:It_as_Ik_when_t_large}\\
    &I_t=I_k+\sum_{i=k}^{H}{(Q_i-\lambda_i)}+\sum_{i=1}^{t-1}\left(Q_i-\lambda_i\right) &\forall t < k \label{eq:It_as_Ik_when_t_small}
\end{align}

Plug in Equation (\ref{eq:It_as_Ik_when_t_large}) and (\ref{eq:It_as_Ik_when_t_small}) into Equation (\ref{eq:benefit_rewrite}), the benefit expression becomes:
\begin{align}
    &\frac{\rho_3}{H} \left[\left(I_k+Q_k-\frac{\lambda_k}{2}\right) + \sum_{t=1}^{k-1}\left(I_t+Q_t-\frac{\lambda_t}{2}\right) + \sum_{t=k+1}^{H}\left(I_t+Q_t-\frac{\lambda_t}{2}\right)\right] \nonumber \\
    =&\frac{\rho_3}{H} \left[ HI_k + Q_k-\frac{\lambda_k}{2} + \sum_{t=1}^{k-1}\left[\sum_{i=k}^{H}\left(Q_i-\lambda_i\right)+\sum_{i=1}^{t-1}{{(Q}_i-\lambda_i)}+Q_t-\frac{\lambda_t}{2}\right] +\sum_{t=k+1}^{H}\left[\sum_{i=k}^{t-1}\left(Q_i-\lambda_i\right)+Q_t-\frac{\lambda_t}{2}\right]\right] \nonumber \\ 
    =&\frac{\rho_3}{H} \left[ HI_k+\sum_{t=1}^{H}Q_t-\sum_{t=1}^{H}\frac{\lambda_t}{2}+\sum_{t=1}^{k-1}\left[\sum_{i=k}^{H}\left(Q_i-\lambda_i\right)+\sum_{i=1}^{t-1}{{(Q}_i-\lambda_i)}\right]+\sum_{t=k+1}^{H}\sum_{i=k}^{t-1}\left(Q_i-\lambda_i\right)\right] \label{eq:benefit_rewritten_no_h_fct} 
\end{align}

Let $h(\bm{Q}, \bm{\lambda})$ be a function of only $\bm{Q}=[Q_1,\ldots,Q_H]$ and $\bm{\lambda}=[\lambda_1,\ldots,\lambda_H]$:
\begin{align}
    &\frac{\rho_3}{H}\sum_{t=1}^{H}\left[I_t+Q_t-\frac{\lambda_t}{2}\right]=\frac{\rho_3}{H} \left[ HI_k+h(\bm{Q}, \bm{\lambda}) \right] \label{eq:benefit_rewritten_all} \\
    &\quad\quad\quad \text{where, } h(\bm{Q}, \bm{\lambda}) = \sum_{t=1}^{H}Q_t-\sum_{t=1}^{H}\frac{\lambda_t}{2}+\sum_{t=1}^{k-1}\left[\sum_{i=k}^{H}\left(Q_i-\lambda_i\right)+\sum_{i=1}^{t-1}{{(Q}_i-\lambda_i)}\right]+\sum_{t=k+1}^{H}\sum_{i=k}^{t-1}\left(Q_i-\lambda_i\right) \nonumber
\end{align}

With Equation (\ref{eq:benefit_rewritten_all}), the objective function can be transformed as:
\begin{align}
    V=&\frac{\rho_3}{H}\left[HI_k+h\left(\bm{Q},\ \bm{\lambda}\right)\right]-\left[\frac{\rho_1}{H}\sum_{t=1}^{H}\left(d_t+\delta_tTT_{i_t,t}\right)+\rho_2I_k+\frac{1}{H}\sum_{t=1}^{H}\left(\delta_t{TC}_{i_t,t}\right)\right] \nonumber \\
    =& \left(\rho_3-\rho_2\right)I_k+\frac{\rho_3}{H}h\left(\bm{Q},\ \bm{\lambda}\right)-\left[\frac{\rho_1}{H}\sum_{t=1}^{H}\left(d_t+\delta_tTT_{i_t,t}\right)+\frac{1}{H}\sum_{t=1}^{H}\left(\delta_t{TC}_{i_t,t}\right)\right] 
\end{align}

Recall that $\rho_2>\rho_3$ for the model to be bounded (Appendix \ref{appendixII_1:rho2_boundproof}), and the {\it Replenish condition} ensures non-negative $I_t\ \forall t$. Hence, the objective function $V$ monotonically decreases with respect to $I_k$, and the optimal solution will always result in the following equation:
\begin{align}
    I_{min}=I_k=0 \label{eq:I_min_zero_property}
\end{align}
    \section{Objective Function Slope Derivation}\label{appendixIII:obj_function_slope}

In this appendix, the slope of the objective function $V$ with respect to the decision variables $d_t \ \forall t\in\{1,\ldots,H\}$ is derived to construct the solution procedure of the {\it Deterministic Model}.

By {\it Conservation of psychological inventory} and {\it Periodicity} constraints, respectively Equation (\ref{eq:dm_cons}) and (\ref{eq:dm_periodicity}), the total production is equal to the total consumption over the planning horizon:
\begin{align}
    \sum_{t=1}^{H}Q_t=\sum_{t=1}^{H}\lambda_t \label{eq:prod_depletion_same_total}
\end{align}

Therefore, Equation (\ref{eq:benefit_rewritten_no_h_fct}) for expressing the benefit expression can be rewritten into: 
\begin{align}
    \frac{\rho_3}{H}\sum_{t=1}^{H}\left[I_t+Q_t-\frac{\lambda_t}{2}\right]
    = &\frac{\rho_3}{H}\left[ HI_k+\sum_{t=1}^{H}\frac{\lambda_t}{2}+\sum_{t=1}^{k-1}\left[\sum_{i=k}^{H}\left(Q_i-\lambda_i\right)+\sum_{i=1}^{t-1}{{(Q}_i-\lambda_i)}\right]+\sum_{t=k+1}^{H}\sum_{i=k}^{t-1}{(Q_i-\lambda_i)} \right] \nonumber \\ 
    = &\frac{\rho_3}{H}\left[ HI_k+\sum_{t=1}^{H}\frac{\lambda_t}{2}+(*) \right] \label{eq:benefit_with_star} \\
    \text{where, }  & (*) = \sum_{t=1}^{k-1}\left[\sum_{i=k}^{H}\left(Q_i-\lambda_i\right)+\sum_{i=1}^{t-1}{{(Q}_i-\lambda_i)}\right]+\sum_{t=k+1}^{H}\sum_{i=k}^{t-1}{(Q_i-\lambda_i)} \nonumber
\end{align}

Equation $(*)$ can be further simplified: 
\begin{align}
    (\ast)=&\sum_{i=k}^{H}\left(Q_i-\lambda_i\right)+\left[\sum_{i=k}^{H}\left(Q_i-\lambda_i\right)+\left(Q_1-\lambda_1\right)\right] + \ldots + \left[\sum_{i=k}^{H}\left(Q_i-\lambda_i\right)+\sum_{i=1}^{k-2}\left(Q_i-\lambda_i\right)\right] \nonumber \\ 
            &+\sum_{i=k}^{k}{(Q_i-\lambda_i)} +\sum_{i=k}^{k+1}{(Q_i-\lambda_i)} +... +\sum_{i=k}^{H-1}{(Q_i-\lambda_i)} \nonumber \\ 
            =& \left(H-1\right)\left(Q_k-\lambda_k\right) \nonumber \\ 
            & +\left[\left(k-1\right)+\left(H-\left(k+1\right)\right)\right]\left(Q_{k+1}-\lambda_{k+1}\right) + \ldots+\left(k-1\right)\left(Q_H-\lambda_H\right) \nonumber \\ 
            & +\left(k-1-1\right)\left(Q_1-\lambda_1\right) +\left(k-1-2\right)\left(Q_2-\lambda_2\right)+\ldots+\left[\left(k-1\right)-\left(k-1\right)\right](Q_{k-1}-\lambda_{k-1}) \nonumber \\
    \Leftrightarrow \quad (\ast) &= \sum_{t=k}^{H}{\left(k-1+H-t\right)(Q_t-\lambda_t)}+\sum_{t=1}^{k-1}\left(k-1-t\right)\left(Q_t-\lambda_t\right) \label{eq:simplifid_ast_eq}
\end{align}

Plug Equation (\ref{eq:simplifid_ast_eq}) into Equation (\ref{eq:benefit_with_star}) for the benefit, and recall that $Q_t=C p_1 d_t$ (Section \ref{section:alt_alg_preliminary}), the objective function is thus: 
\begin{align}
    V =&\frac{\rho_3}{H}\left[ HI_k+\sum_{t=1}^{H}\frac{\lambda_t}{2}+\sum_{t=k}^{H}{\left(k-1+H-t\right)(C p_1 d_t-\lambda_t)}+\sum_{t=1}^{k-1}\left(k-1-t\right)\left(C p_1 d_t-\lambda_t\right) \right] \nonumber \\
        &-\left[\frac{\rho_1}{H}\sum_{t=1}^{H}\left(d_t+\delta_tTT_{i_t,t}\right)+\rho_2I_k+\frac{1}{H}\sum_{t=1}^{H}\left(\delta_t{TC}_{i_t,t}\right)\right] \label{eq:reformed_obj_fun}
\end{align}

Now, the slope of the objective function with respect to duration can be derived as shown below, which is the same form as Equation (\ref{eq:slope_eq}):
\begin{align}
    \Delta_t = \frac{\partial V}{\partial d_t} = \begin{cases}
        \left(k-1+H-t\right)Cp_1\frac{\rho_3}{H}-\frac{\rho_1}{H}\ & t\geq\ k \\
        \left(k-1-t\right)Cp_1\frac{\rho_3}{H}-\frac{\rho_1}{H}\ & t< k
    \end{cases} \nonumber
\end{align}

\end{APPENDICES}

\bibliographystyle{informs2014trsc} %
\bibliography{References} %

\end{document}